\documentclass[aos,preprint]{imsart}

\RequirePackage{amsthm,amsmath,amsfonts,amssymb}
\RequirePackage[numbers]{natbib}
\RequirePackage[colorlinks,citecolor=blue,urlcolor=blue]{hyperref}
\RequirePackage{graphicx}

\usepackage{textcomp,enumerate,bbm,latexsym, bm,xcolor}
\usepackage{url,algorithm,algorithmic,verbatim}
\usepackage{multirow}
\usepackage{booktabs}
\usepackage{cleveref}
\usepackage{chngcntr, subcaption}
\usepackage{bm}
\usepackage{dsfont}

\def\spacingset#1{\renewcommand{\baselinestretch}%
{#1}\small\normalsize} \spacingset{1}

\startlocaldefs

\newtheorem{theorem}{Theorem}[section]
\newtheorem{lemma}[theorem]{Lemma}
\newtheorem{cor}[theorem]{Corollary}
\newtheorem{prop}[theorem]{Proposition}
\theoremstyle{remark}
\newtheorem{definition}[theorem]{Definition}

\newtheorem{assumption}{Assumption}
\newtheorem{rem}{Remark}[section]
\allowdisplaybreaks  

\newcommand{\real}{\mathbb{R}}

\renewcommand{\Pr}{\mathbb{P}}
\newcommand{\e}{\mathbb{E}}

\newenvironment{enumeratealpha}{\begin{enumerate}[(a)] }{\end{enumerate}}

\endlocaldefs

\begin{document}

\begin{frontmatter}
\title{Detecting Multiple Replicating Signals using Adaptive Filtering Procedures}
\runtitle{AdaFilter for Replicating Signals}

%
%

\author{Jingshu Wang$^{1,*}$
	Lin Gui$^{1}$,
	Weijie J. Su$^{2}$,
	Chiara Sabatti$^{3}$, and Art B. Owen$^{3}$ \\
	$^{1}$Department of Statistics, University of Chicago. \\
	$^{2}$Department of Statistics and Data Science, University of Pennsylvania. \\
	$^{3}$Department of Statistics, Stanford University}

\begin{abstract}  
Replicability is a fundamental quality of scientific discoveries: we are interested in those signals that are detectable in different laboratories, different populations, across time etc.
Unlike meta-analysis which accounts for experimental variability but does not guarantee replicability, testing a partial conjunction (PC) null aims specifically to identify the signals that are discovered in multiple studies. 
In many contemporary applications, 
e.g., comparing multiple high-throughput genetic experiments, a large number $M$ of PC nulls need to be tested simultaneously, calling for a multiple comparisons correction.
However, standard multiple testing adjustments on the $M$ PC $p$-values can be severely conservative, especially when $M$ is large and the signals are sparse. 
We introduce AdaFilter, a new multiple testing procedure that increases power by adaptively filtering out unlikely candidates of PC nulls. 
 We prove that AdaFilter can control FWER and FDR as long as data across studies are independent, and has much higher power than other existing methods.  
 We illustrate the application of AdaFilter with three examples: microarray studies of Duchenne muscular dystrophy, single-cell RNA sequencing of T cells in lung cancer tumors and GWAS for metabolomics.
\end{abstract}



\end{frontmatter}




\section{Introduction}\label{sec:intro}

Replication is ``the cornerstone of science" \citep{moonesinghe2007}. 
An important scientific finding should be supported by further evidence from similar conditions, by other researchers or with new samples\nocite{telvivpage}.
In the last decade,  however, both the  popular  \citep{NY} and the scientific press \citep{amgen, baker2016} have 
reported the lack of replicability in modern research.
While there are many reasons behind this phenomenon,  one important factor is that many scientific discoveries are obtained from complicated large-scale experiments with biases from various sources. 
Even when the data are carefully analyzed, idiosyncratic aspects of a single experiment can fail to extend to other settings, and any finding from just one study can easily lack external validity. 
Thus, it is crucial to have a statistical framework  to objectively and precisely evaluate the consistency of scientific discoveries across multiple studies, while properly accounting for experimental heterogeneity.

The partial conjunction (PC) test, which was 
introduced by  \cite{fris:penn:glas:2005} and further studied in \cite{benjamini2008}, provides such a framework. Given $n$ null hypotheses (base nulls) and a number $r\in\{2,3,\dots,n\}$, the PC null states that there are fewer than $r$  base non-nulls. 
In the setting where each base hypothesis represents a test from one study, rejecting a PC null explicitly guarantees that the signal replicates at least $r$ times. The PC framework has been used to identify replicating signals in neuroimaging \citep{pric:fris:1997}, to detect genes that show consistent effects across genetic experiments \citep{heller2014deciding}, and recently to study mediation effects \citep{liu2021large} and find evidence factors \citep{karmakar2021reinforced} in causal inference.  

In high-throughput genetic experiments, there is a special need to identify replicating signals across multiple studies. For instance, for gene expression data, it is important to find stable gene markers for a disease or cell type, which remain differentially expressed across similar experiments or in multiple patients. In multi-tissue expression quantitative trait loci (eQTL) studies, scientists are interested in identifying DNA loci with  consistent  regulation  over tissues \cite{flutre2013,urbut2019flexible}. With a growing trend in multi-omics data sharing \citep{hasin2017multi}, there is also active research in finding replicating signals across platforms \citep{zhang2010joint}, ethnic groups \citep{marigorta2013high,giri2019trans}  and even species. 
Though the PC framework fits all above scenarios, finding multiple replicating signals by simultaneously performing a large number of PC tests for thousands of genes or millions of DNA loci, however, typically suffers from extremely low power.

Specifically, let $M$ denote the number of hypotheses in one study and suppose that we compare across $n$ related studies. Then, to find replicating signals across the $n$ studies, we have $M$ PC nulls to test, each with $n$ base nulls.
The above framework gives us an $n\times M$ matrix of base $p$-values, with one column per PC null and one row per study. 
Now, as we want to identify signals whose PC nulls are false, a ``direct approach'' is to first get a combined $p$-value for each PC null and then apply standard multiple testing adjustment to the $M$ PC p-values.
However, 
this ``direct approach'' for testing multiple PC tests has been shown to have extremely low power \citep{heller2014,sun2015}. 
Both  \cite{heller2014} and \cite{bogomolov2018assessing} suggest procedures to counter that power loss. Unfortunately, the appoach in  \cite{bogomolov2018assessing} is designed only for $n = r = 2$ and the empirical Bayes approach {\tt repfdr} in \cite{heller2014} encounters both accuracy and computational barriers for $n$ as large as $8$, as shown in our simulations. There is thus a need for a powerful and fast method that can guarantee simultaneous error control and can handle a larger number of studies.

In this paper, we introduce AdaFilter, an adaptive filtering multiple testing procedure for multiple PC hypotheses. 
We propose different versions of AdaFilter to control simultaneous error rates including FDR (false discovery rate) and FWER (familywise error rate). AdaFilter can control FWER and FDR when all $nM$ base p-values are independent. In addition, it asymptotically controls FDR when $M$ goes to infinity, allowing base p-values to be weakly associated within each study. The weak dependence only assumes that within each study, the number of pairs $(j, j')$ where the base p-values $p_j$ and $p_{j'}$ are dependent is $o(M^2)$, which is reasonable for most genetics and genomics data. 
Using simulations and real data applications, we show that AdaFilter is robust to dependence of p-values within each study and can have much higher power than the ``direct approach'' or using {\tt repfdr}.

Deferring precise statements to later sections, we give an intuitive explanation for how AdaFilter gains power. The low power of the ``direct approach'' is due to the fact that partial conjunction has a composite null. 
AdaFilter's power gain is linked to its ability to borrow information across studies and learn from the data which PC hypotheses are likely to be least favorable nulls.
Intuitively, AdaFilter filters the  set of hypotheses down to a number $m<M$ of candidate least favorable nulls, which are the nulls that have exactly $r -1$ base non-nulls. The PC p-values are still ``valid'' conditioning on filtering and the decreased number of hypotheses lowers multiplicity burden. More surprisingly, the power gain also links to a lack of ``monotonicity'' of the  number rejections in the base p-values, where increasing some base p-values can result in more rejections. In the extreme case, combining multiple studies while requiring replicability can even lead to more rejections than the union of rejections by testing each individual study separately.

The structure of the paper is as follows. \Cref{sec:pre} precisely defines the PC framework, and illustrates the power limitation of the ``direct approach''. 
\Cref{sec:def} introduces our AdaFilter procedures. \Cref{sec:theory} discusses theoretical properties of AdaFilter. \Cref{sec:simu} explores the performance with simulations. \Cref{sec:real} applies AdaFilter to several real studies. 
\Cref{sec:conc} has conclusions. An R package implementing AdaFilter is available at \url{https://github.com/jingshuw/adaFilter}.

\section{Multiple testing for partial conjunctions}\label{sec:pre}

In this section, we provide a brief introduction of the partial conjunction hypotheses and the low power in detecting multiple PC hypotheses using the ``direct approach''.

\subsection{Problem setup}\label{sec:setup}

We consider the problem where $M$ null hypotheses are tested in $n$ studies. The base null hypotheses are $(H_{0ij})_{n \times M}$. In high-throughput experiments, $M$ is the number of genes or DNA loci. 
We work with summary statistics that are base p-values  $(p_{ij})_{n \times M}$ for $(H_{0ij})_{n \times M}$. Each $p_{ij}$ is the realization of a random variable $P_{ij}$. We assume that each base P-value is valid, satisfying 
$\Pr(P_{ij}\le\gamma)\le\gamma$ under its null. 
Also, let $P_{(1)j} \leq P_{(2)j} \leq \cdots \leq P_{(n)j}$ be the sorted P-values for each $j = 1, 2, \ldots, M$.

\begin{definition}[Partial Conjunction Hypothesis]\label{defn:pch0}
For integers $n\ge r\ge2$, the partial conjunction (PC) null hypothesis is:
$$H_{0}^{r/n}: \mathrm{fewer\ than\ } r \mathrm{\ out\ of\ } n \mathrm{\ base\ hypotheses\ are\ \text{non-null}}.$$ 
\end{definition}

When $r = 1$, $H_0^{1/n}$ is the commonly tested global null for meta-analysis. Rejecting it would not guarantee replicability. 
In high-throughput experiments, for each DNA locus or gene $j \in \{1, 2, \ldots, M\}$, we test for a PC null $H_{0j}^{r/n}$ to evaluate if genetic signals have been replicated at least $r$ times across $n$ studies. Throughout the paper, we assume that p-values across studies are independent. This can be assumed when samples do not overlap across studies.

For a multiple testing procedure on $\{H_{01}^{r/n}, \ldots, H_{0M}^{r/n}\}$, denote the decision function as $\varphi_j = 1$ if we reject $H_{0j}^{r/n}$ and $\varphi_j = 0$ otherwise. The total number of discoveries is then $R = \sum_{j = 1}^M \varphi_j$. Among these, the number of false discoveries is $V = \sum_{j = 1}^M \varphi_j1_{v_j = 0}$ where $v_j = 0$  if $H_{0j}^{r/n}$ is true and $v_j = 1$ otherwise.

There are many measures of the simultaneous error rate \citep{dudoit2007}, with FWER and FDR being the most common ones.
In addition, we consider the per-family error rate (PFER), as it provides a motivation for our procedures. With the notation introduced, we have
$$\text{FWER}:=\Pr(V\ge1), \quad
\text{PFER}:=\e(V), \quad \text{FDR}:=\e( \text{FDP}).$$ 
where $\text{FDP} = V/(R\vee 1)$ is the false discovery proportion.

\subsection{The ``direct approach''}\label{sec:classical}

We start with a brief review of p-value construction for a single PC null, while more details can be found in \cite{wang2018admissibility} and \cite{benjamini2008}. Consider a single PC null $H_0^{r/n}$ with a vector of base P-values $(P_1, P_2, \ldots, P_n)$ and let $P_{r/n} = f(P_1, P_2, \ldots, P_n)$ be the combined P-value for $H_{0}^{r/n}$. Benjamini and Heller \cite{benjamini2008} 
discussed three approaches, which we report here, using the standard notation $(P_{(1)} \leq P_{(2)} \leq \cdots\leq  P_{(n)})$:
\begin{enumerate}
    \item Simes' method:
$$P_{r/n}^S = 
\min_{\,r\le i\le n} \Bigl\{ \frac{n - r + 1}{i - r + 1}P_{(i)}  \Bigr\},$$
\item Fisher's method:
$$P_{r/n}^F = \Pr \bigl( \chi_{(2(n -r + 1))}^2 \geq - 2 
\sum_{i = r}^n \log P_{(i)}),$$
\item Bonferroni's method:
      $$P_{r/n}^B =  (n - r + 1)P_{(r)}.$$
\end{enumerate}
The idea is to apply meta-analysis to the largest $n -r + 1$ base P-values. 
For instance, if $n = r = 2$, then $P_{2/2}^S = P_{2/2}^F = P_{2/2}^B = \text{max}(p_1, p_2)$. 
All three methods construct valid PC $P$-values for $H_{0}^{r/n}$ under independence, and \cite{wang2018admissibility} showed that they also provide the most powerful tests for a single PC null. For $M$ hypotheses, we denote $P_{r/n, j}$ as the PC p-value for the $j$th PC null.

The ``direct approach'' is to simply apply standard multiple testing adjustment procedures to the $M$ PC P-values.  For example, to control the FWER at level $\alpha$, we could use the 
Bonferroni rule, rejecting $H^{r/n}_{0j}$ 
if $P_{r/n, j}\leq \alpha/M$, which also controls the PFER at level $\alpha$ \citep{tukey1953}.
To control the FDR we could apply BH procedure \citep{benjamini1995} on $\{P_{r/n, j}, j = 1, \cdots, M\}$.
 
However, this direct approach is often too conservative, as we illustrate now for the case $r=n$.
To quantify how the performance associates with the composite nature of a PC null,
define sets $\mathcal{I}_k\subset \{1, \cdots, M\}$ such that 
\begin{equation}\label{eq:partition}
    \mathcal{I}_k= \big\{ j\in \{1, \cdots, M\} \mid \text{exactly $k$ of $H_{01j},\ldots,H_{0nj}$ are false}\big\} 
\end{equation}
for $k=0,\dots,n$. Sets $\{\mathcal{I}_k$, $k = 0,\dots,n\}$ define a partition of $\{1,\dots,M\}$.
If a false rejection of $H_{0j}^{n/n}$ happens, then the $j$th column must belong to one of $\mathcal{I}_k$ where $k = 0, 1, \cdots, n-1$. Thus, if we use Bonferroni to control for FWER at a nominal level $\alpha$, the true FWER instead satisfies
\begin{align*}
 \text{FWER} \leq \e(V) 
 & = \sum_{k=0}^{n-1}\sum_{j\in{\cal I}_k} \Pr (P_{(n)j} \leq \alpha/M) \\
 & \leq \sum_{k = 0}^{n-1}\sum_{j\in{\cal I}_k} \frac{\alpha^{n -k}}{M^{n -k}}
  = \sum_{k = 0}^{n-1} |\mathcal{I}_k| \frac{\alpha^{n -k}}{M^{n -k}}.
 \end{align*}
where the second inequality is close to an equality when 
all the tests for non-nulls $H_{1ij}$ have high power. 
Let $\delta_k = |\mathcal{I}_k|/M$ be the proportion of hypotheses in each partition. Then 
we have 
\begin{equation}\label{eq:bound}
    \e(V)\leq  \alpha\Bigl\{\delta_{n-1} + \delta_{n-2}\frac{\alpha}M + \delta_{n-3}\Bigl(\frac{\alpha}M\Bigr)^2 +\cdots+\delta_0\Bigl(\frac\alpha{M}\Bigr)^{n-1}\Bigr\}
\end{equation}
which in the limit is dominated by $\delta_{n-1}\alpha$ (when
$\delta_{n-1} \neq 0$) 
or is of order $O(M^{-1})$ (when $\delta_{n-1} = 0$) for large $M$. 
Thus, when $\delta_{n - 1} \approx 0$, a typical scenario in genetics problems with sparse signal, 
the expected number of rejections $\e(V)$ would be much smaller than $\alpha$ and the ``direct approach'' can become highly deficient, 
in fact much more conservative than Bonferroni usually is.

 The point is that  if we do not 
account for the fact that the PC null is composite, we will control the simultaneous error rates under the worst case scenario  ($\delta_{n-1}=1$), which is unnecessary.  
For general $r \leq n$, 
the level of $\e(V)$ for Bonferroni correction
will depend mainly on $\delta_{r - 1}$ in the large $M$ setting. So does the BH control for FDR.

It is clear that there can be more efficient procedures if the fractions $\delta_k$ were known 
or if good estimates of $\delta_k$ can be obtained. 
This is what motivates the Bayesian methods \citep{heller2014,flutre2013}.  In this paper we take a frequentist perspective. Rather than estimating $\delta_k$,  AdaFilter works directly on an alternative estimation of $V$,  implicitly and adaptively adjusting for the size of $\delta_{r - 1}$, the fraction of the least favorable nulls.

\section{The idea of AdaFilter}\label{sec:def}

In \Cref{sec:classical}, we showed that a PC null hypothesis is  composite, thus the inequality $\Pr( P_{r/n}\leq\gamma)\leq\gamma$ for a given $\gamma$ is only tight for the least favorable null, while 
standard multiple testing procedures are designed to control error when $\Pr( P_{r/n}\le\gamma)=\gamma$ is always true. To overcome this, AdaFilter leverages 
a region $\mathcal{A}_\gamma \subset[0,1]^n$ such that the much tighter inequality 
\begin{equation*}
    \Pr(P_{r/n, j} \leq \gamma \mid (P_{1j}, \ldots, P_{nj}) \in \mathcal{A}_\gamma) \leq \gamma
\end{equation*}
holds for any configuration  in the PC null space.

\begin{figure}[ht]
  \center 
 \includegraphics[width =0.55\textwidth]{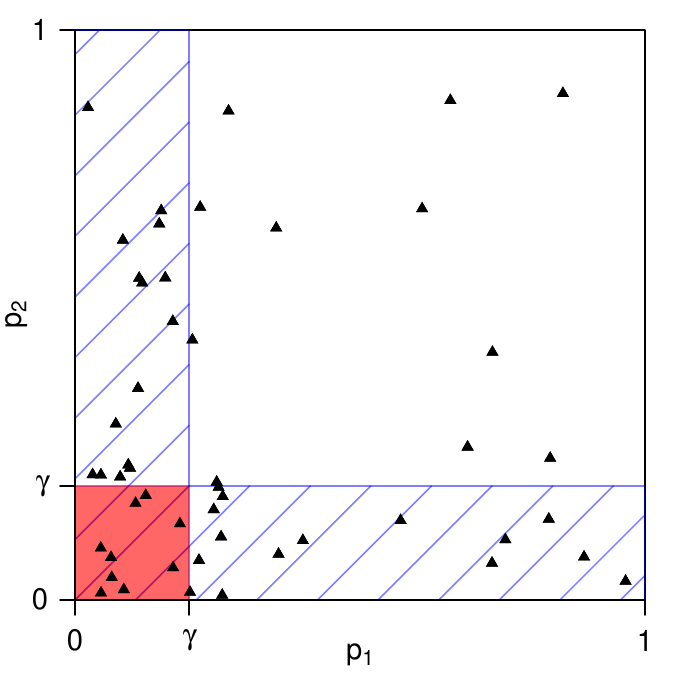}
\caption{Illustration of the rejection (red) and filtering (L-shaped blue) regions at $\gamma = 0.2$ when $n = r = 2$. Each triangle corresponds to a pair of p-values. 
}
\label{fig:demo}
\end{figure}

\Cref{fig:demo} illustrates the construction of the filtering region $\mathcal{A}_\gamma$ for $r = n = 2$. The PC test $j$ has base $p$-values $P_{1j}$ and $P_{2j}$, and its PC $p$-value is $P_{2/2, j} = \max(P_{1j}, P_{2j})$. 
The null $H^{2/2}_{j0}$ contains three configurations: $(H_{01j},H_{02j})$ being (True, True), (True, False) or (False, True). It is easy to see that $\Pr(P_{2/2, j}\leq \gamma)\leq\gamma^2$ under (True, True), while $\Pr(P_{2/2, j}\leq \gamma)$ can be close to $\gamma$ under the other two less favorable configuration. 
Let us consider, instead, conditioning on $(P_{1j},P_{2j})$ being in the ``L''-shaped filtering region $\mathcal{A}_\gamma = \{(p_1, p_2)\mid \min(p_1, p_2) \leq \gamma\}$.
We get 
$\Pr(P_{2/2, j} \leq \gamma \mid (P_{1j}, P_{2j}) \in \mathcal{A}_\gamma) \leq \gamma$ being true for all three null scenarios, which is much tighter than $\Pr( P_{r/n}\leq\gamma)\leq\gamma$. The inequality holds since at least one of $P_{1j}$ and $P_{2j}$ is stochastically greater  than uniform under all three configurations. 

Since  Bonferroni and BH procedures are based on an implicit estimate of the number of false rejections $V$ associated with a threshold $\gamma$: $\widehat V_{\gamma} = \gamma M$, we can improve their efficiency with a smaller estimate of $\widehat V_{\gamma}$ using the new inequality. 
Specifically, the estimated $V$ is now $\widehat V_{{\cal A}_{\gamma}} = \gamma \times \sum_{j = 1}^ M1_{(P_{1j}, P_{2j}) \in \mathcal{A}_\gamma}$, where $M$ is replaced by the number of hypotheses falling into the $L$ shaped region, a possibly much smaller number than $M$. Alternatively, the quantity $(1/M)\sum_{j = 1}^ M1_{(P_{1j}, P_{2j}) \in \mathcal{A}_\gamma}$ is our ``estimate'' of $\delta_{r-1}$, the fraction of least favorable nulls. 
Hypotheses that fall outside of the ``L''-shaped filtering region are not counted towards the multiplicity of the PC hypotheses.

To control the FWER (and PFER) at level $\alpha$, 
we can adaptively choose the largest $\gamma$ satisfying $\widehat V_{{\cal A}_{\gamma}}\leq \alpha$. 
Similarly, to control the FDR at level $\alpha$, we estimate the FDP as $\widehat V_{{\cal A}_{\gamma}} /(R \vee 1)$ and select the largest $\gamma$ such that $\widehat V_{{\cal A}_{\gamma}} /(R\vee 1) \leq \alpha$. These are essentially the Bonferroni or BH procedure with an alternative estimate of $V$.

\subsection{Definition of AdaFilter procedures}

Now we formally define AdaFilter for general $n$ and $r$.
It is convenient to first introduce the notion of {\em filtering} and {\em selection} ``$P$-values''. These are
 \begin{align} F_j &:= (n -r + 1)P_{(r-1)j},\label{eq:filterp}\quad\text{and}\\
 S_j &:= P_{r/n, j}^B = (n -r + 1)P_{(r)j}, \label{eq:selectp} 
 \end{align} 
respectively. 

\begin{definition}[AdaFilter Bonferroni]\label{def:two_step_MTP_bonf}
For a  level $\alpha$, and with $F_j$ and $S_j$ given
by \eqref{eq:filterp} and~\eqref{eq:selectp} respectively, 
reject $H_{0j}^{r/n}$ if $S_j< \gamma_0^{\text{Bon}}$ where 
\[
  \gamma_0^{\text{Bon}} = \sup\Big\{\gamma\in[0,\alpha]
     ~\Big |~ 
\gamma {\sum_{j = 1}^M 1_{F_j < \gamma}}\leq \alpha \Big\}.
\]

\end{definition}
\begin{definition}[AdaFilter BH]\label{def:two_step_MTP_BH}
For a  level $\alpha$, and with $F_j$ and $S_j$ given
by \eqref{eq:filterp} and~\eqref{eq:selectp} respectively,
    reject $H_{0j}^{r/n}$ if $S_j < \gamma_0^{\text{BH}}$ where 
 \begin{align*}
  \gamma_0^{\text{BH}} &= \sup\left\{\gamma \in [0,\alpha] ~\Big |~ 
\frac{\gamma {\sum_{j = 1}^M 1_{F_j < \gamma}}}{{\sum_{j = 1}^M 1_{S_j < \gamma}} \vee 1}\leq \alpha 
\right\}.
\end{align*}
\end{definition}

\begin{rem}
We define the filtering region as $\{F_j < \gamma\}$ instead of $\{F_j \leq \gamma\}$ to guarantee that $\gamma_0^{\text{Bon}}$ and $\gamma_0^{\text{BH}}$ themselves satisfy the corresponding inequalities. This is important for showing the theoretical properties of adaFilter procedures, especially when base p-values are discrete. The rejection criterion is set to $ S_j < \gamma_0$ instead of $S_j \leq \gamma_0$ where $\gamma_0$ is either $\gamma_0^{\text{Bon}}$ or $\gamma_0^{\text{BH}}$ accordingly (for \Cref{lem:frac}). 
\end{rem}

We also introduce AdaFilter adjusted ``p-values'' like those commonly computed for standard Bonferroni and BH procedures. They provide equivalent sets of rejections as the above definitions, while can be more efficiently computed. 

\begin{definition} [AdaFilter adjusted p-values]\label{def:two_step_MTP_equiv}
Rank the selection p-values as $S_{(1)} \leq S_{(2)} \leq \cdots \leq S_{(M)}$ where $S_{(j)}$ is for the null hypothesis $H_{0(j)}^{r/n}$. 
For each $j$, define an AdaFilter adjustment number 
$$m_{(j)}^\text{AF} := \sum_{h = 1}^M 1_{F_h \le S_{(j)}}.$$
Then the AdaFilter Bonferroni adjusted P-value for $H_{0(j)}^{r/n}$ is
$$P_{(j)}^{\text{Bon}} =S_{(j)}m_{(j)}^\text{AF}$$
and the AdaFilter BH adjusted P-value for $H_{0(j)}^{r/n}$ is 
$$P_{(j)}^{\text{BH}} = \min\left\{\min_{h \geq j}\left\{S_{(h)}\frac{m_{(h)}^\text{AF}}{h}\right\}, 1\right\}.$$
\end{definition}

For any level $\alpha>0$, we reject the hypotheses whose AdaFilter adjusted p-values are smaller than $\alpha$. We can verify that the AdaFilter adjusted p-values give the same set of rejections as \Cref{def:two_step_MTP_bonf} and \Cref{def:two_step_MTP_BH}.

 \begin{prop}\label{prop:equiv}
   For any level $\alpha > 0$, the set of rejections defined as $\{j: P_j^{\text{Bon}} < \alpha\}$ is equivalent to the set of rejections from \Cref{def:two_step_MTP_bonf}. Similarly, the set of rejections defined as $\{j: P_j^{\text{BH}} < \alpha\}$ is equivalent to the set of rejections from \Cref{def:two_step_MTP_BH}. 
  \end{prop}
In practice, the AdaFilter adjusted p-values can be more easily computed than finding $\gamma_0^{\text{Bon}}$ and $\gamma_0^{\text{BH}}$. Our simulations and real data applications in \Cref{sec:simu,sec:real} also compute these adjusted p-values for getting the rejections of AdaFilter procedures.

\subsection{A heuristic comparison with the ``direct approach''}
Before we discuss the theoretical properties of AdaFilter procedures in Section~\ref{sec:theory}, we revisit the case of  $r=n$ in Section~\ref{sec:classical} to understand the level of power gain from AdaFilter procedures compared with the ``direct approach''. 
When $r = n$, the PC p-values for the ``direct approach'' are $P_{r/n,j} = P_{(n)j}$, which are the same as the selection p-values of AdaFilter procedures. As a consequence, AdaFilter procedures would not change the ordering/ranking of the individual PC hypotheses. AdaFilter gains power by selecting a much less conservative PC p-values threshold $\gamma$ than the ``direct approach'' for the same nominal FWER/FDR level. 

If one controls FWER at level $\alpha$, then the PC p-value threshold from the ``direct approach'' using Bonferroni adjustment is $\alpha/M$. We now give an approximation of the threshold from AdaFilter Bonferroni. 
When $r = n$, at any given threshold $\gamma$, the estimate of the number of false discoveries used in AdaFilter is 
$$\hat{V}(\gamma)=\gamma\sum_{i=1}^M1_{F_j<\gamma}=\gamma\sum_{i=1}^M1_{P_{(n-1),j}<\gamma}.$$
AdaFilter Bonferroni finds the largest $\gamma$ so that $\hat{V}(\gamma) \leq \alpha$. As defined in \eqref{eq:partition}, let $\mathcal{I}_k\subset \{1, \cdots, M\}$ be the set of hypotheses with exactly $k$ base non-nulls and let $\delta_k = |\mathcal{I}_k|/M$. When $M$ is large, the expected value of $\hat{V}(\gamma)$ satisfies that 
\begin{align*}
\mathbb{E}\left(\hat{V}(\gamma)\right)&=\gamma\sum_{k=0}^n\sum_{j\in{\cal I}_k}\mathbb{P}\left(P_{(n-1),j}<\gamma\right)\\
&\le\gamma\left(|\mathcal{I}_n|+|\mathcal{I}_{n-1}|+\sum_{k=0}^{n-2}|\mathcal{I}_k|\cdot\left((n-k)\gamma^{n-k-1}(1-\gamma)+\gamma^{n-k}\right)\right)\\
&\le \gamma M(\delta_n+\delta_{n-1})+ M O(\gamma^2). 
\end{align*}
The first inequality is due to the fact that all base null p-values are independent and for each $j\in{\cal I}_k$, we can decompose $\mathbb{P}\left(P_{(n-1),j}<\gamma\right)$ into the events that all $n - k$ base nulls $i$ satisfy $P_{ij} \leq \gamma$ and exactly $n - k - 1$ base nulls satisfy this constraint. 
So roughly, the AdaFilter Bonferroni threshold $\gamma^{\text{Bon}}$ will be around some value that is at least $\alpha/\big(M(\delta_{n} + \delta_{n-1})\big) + o(1/M)$. Compared with the Bonferroni threshold $\alpha/M$ in the ``direct approach'', AdaFilter Bonferroni increases this threshold by $1/(\delta_n + \delta_{n - 1})$. In our motivating applications, both $\delta_n$ and $\delta_{n-1}$ are typically small, and so such an increase would be substantial. 
The resulting actual FWER is also less conservative. If we use a fixed threshold at $\gamma = \alpha/\big(M(\delta_{n} + \delta_{n-1})\big)$, then
$$
\e(V) \leq \alpha\Bigl\{\frac{\delta_{n-1}}{\delta_{n-1} + \delta_n} + O(\frac{1}{M})\Bigr\}.
$$
Compared to the bound  $\alpha\delta_{n - 1} + O(1/M)$ in \eqref{eq:bound} from the ``direct approach'', we can now be much less conservative especially when the proportion of least favorable PC nulls $\delta_{n-1}$ is small.

\section{Theoretical properties of AdaFilter}\label{sec:theory}

Now we prove that AdaFilter procedures control simultaneous error rates under various conditions. As stated in \Cref{sec:setup}, all the following results assume that p-values across $n$ studies are independent. The key property that AdaFilter relies on is the following conditional validity lemma:
\begin{lemma}[Conditional validity]\label{lem:frac}
When $H_{0j}^{r/n}$ is true, for any fixed $\gamma > 0$
\begin{equation}\label{eqn:frac}
\Pr\big(S_j < \gamma\mid F_j< \gamma\big)\leq \gamma
\end{equation}
holds whenever  $\Pr\big(F_j<\gamma\big) > 0$. 
Here $F_j$ and $S_j$ are given 
by \eqref{eq:filterp} and~\eqref{eq:selectp}, respectively.
\end{lemma}
Inequality \eqref{eqn:frac} can be equivalently written as 
$\Pr\big(S_j < \gamma\big)\leq \gamma\Pr(F_j < \gamma)$, which holds even when $\Pr(F_j < \gamma) = 0$ as $S_j \geq F_j$ is always true. Intuitively, the ``conditional validity'' guarantees that for a fixed threshold $\gamma$, the estimated upper bound on the number of false rejections $V$ is $\gamma\sum_j 1_{F_j < \gamma}$. However, AdaFilter uses a data-dependent $\gamma$, so extra assumptions on the base p-values within one study are needed to prove simultaneous error control of AdaFilter.

\subsection{Exact simultaneous error rates control for finite $M$}
First, for a finite number of hypotheses $M$, we can show that AdaFilter Bonferroni controls FWER and PFER if we further assume independence of all $nM$ base p-values.

\begin{theorem}\label{thm:validity_two_step}
Let $(P_{ij})_{n \times M}$ contain independent valid $p$-values.
Then AdaFilter Bonferroni in \Cref{def:two_step_MTP_bonf}
controls FWER and PFER at level $\alpha$
for the null hypotheses 
$\{H_{0j}^{r/n}: j = 1, 2, \ldots, M\}$.
\end{theorem}

\begin{rem}
Though we name our method AdaFilter Bonferroni, we can only prove FWER/PFER control under independence of the p-values within each study, though simulations in \Cref{sec:simu} show that FWER/PFER control can also be achieved in practice for dependent p-values within each study. 
\end{rem}

\begin{rem}
For controlling for FWER, one can combine adaFilter Bonferroni with the sequential rejection principle \citep{goeman2010sequential} to further increase the number of rejections while controlling for FWER at the same level. Intuitively, this is similar to improving the standard Bonferroni procedure with Holm's procedure. For a more detailed discussion, see Section S1. 
\end{rem}

For AdaFilter BH, however, we can only prove that it controls FDR at the nominal level of $\alpha C(M)$ where $C(M) = \sum_{j = 1}^M 1/j \approx \log M$. 
In other words, adjusting the threshold to be $\alpha/{C(M)}$ can guarantee control of the FDR at level $\alpha$.
\begin{theorem}\label{thm:validity_FS_BH}
Let $(P_{ij})_{n\times M}$ contain independent valid $p$-values.
Then AdaFilter BH in \Cref{def:two_step_MTP_BH}
controls FDR at level $\alpha C(M)$ where $C(M) = \sum_{j = 1}^M 1/j$
for the null hypotheses $\{H_{0j}^{r/n}: j = 1, 2, \cdots, M\}$.
\end{theorem}

The inflation factor $C(M)$ in \Cref{thm:validity_FS_BH} for the adaFilter BH procedure is due to a technical difficulty encountered when proving for FDR control for finite $M$. In \Cref{sec:simu}, we find in simulations that the AdaFilter BH procedure adjusted by $C(M)$ still achieves higher power than other bench-marking approaches. Our simulations also suggest that the adjustment $C(M)$ is actually not needed in practice. In \Cref{sec:asym}, we will show that AdaFilter BH can asymptotically controls FDR without using the inflation factor $C(M)$ when $M \to \infty$. The asymptotic results also do not require independence among p-values within each study.

\subsection{Asymptotic FDR control when $M \to \infty$} \label{sec:asym}
Now we discuss FDR control of AdaFilter BH when the number of hypotheses $M$ is very large, the usual case in high-throughput genetic experiments.  Inspired by \cite{ferreira2006benjamini}, we make the following three assumptions.

First, instead of requiring independent p-values within each study,  we only assume a weak dependence structure among the p-values within each study. 
\begin{assumption}[Weak dependence]\label{assp:weak}
Within any study $i$, the p-values $P_{ij}$ for $j = 1, 2, \cdots, M$ satisfy weak dependence where for any fixed $\gamma$
$$\frac{1}{M^2}\sum_{j\neq j'}\big|\Pr(P_{ij}<\gamma,P_{ij'}<\gamma)- \Pr(P_{ij}<\gamma)\Pr(P_{ij'}<\gamma)\big|\rightarrow0$$
as $M\rightarrow\infty$.
\end{assumption}

One scenario where the weak dependence holds is that,  within each study $i$, the number of pairs $(P_{ij}, P_{ij'})$ where $P_{ij}$ and $P_{ij'}$ are not independent is $o(M^2)$. For microarrays or RNA-seq experiments, gene-gene networks are typically sparser than $O(M^2)$. For GWAS or eQTLs, DNA loci are usually associated only when they are close enough along 
the DNA chain, say when $|j - j'| < b$ for some constant $b$. The weak dependence assumption is reasonable for both the above two scenarios.

Now let $\mathcal{H}_0^{r/n} = \{j: H_{0j}^{r/n} \text{ is true}\}$ be the set of true PC nulls and $M_0$ be its cardinality. Similarly, define $\mathcal{H}_1^{r/n} = \{j: H_{1j}^{r/n} \text{ is true}\}$ to be the set of true PC non-nulls and let $M_1$ be its cardinality. Besides weak dependence, we also assume that when $M \to \infty$, the following limits exist:

\begin{assumption}[Existence of limits]\label{assp:asymptotic} 
The following limits exist:
$$\lim_{M \to \infty} \frac{M_0}{M} = \pi_0 \in (0, 1)$$
$$\lim_{M\to\infty} \frac{1}{M_0}\sum_{j\in\mathcal{H}_0^{r/n}}P(F_j<\gamma) = \tilde F_0(\gamma) , \quad
\lim_{M\to\infty}\frac{1}{M_1}\sum_{j\in\mathcal{H}_1^{r/n}}P(F_j<\gamma)  = \tilde F_1(\gamma) $$
$$\lim_{M\to\infty}\frac{1}{M_0}\sum_{j\in\mathcal{H}_0^{r/n}}P(S_j<\gamma) =\tilde S_0(\gamma), \quad
\lim_{M\to\infty}\frac{1}{M_1}\sum_{j\in\mathcal{H}_1^{r/n}}P(S_j<\gamma)=\tilde S_1(\gamma). $$
\end{assumption}

For a given $n$, there are $2^n$ combinations of base hypotheses being null or non-null.
A special case where \Cref{assp:asymptotic} is satisfied is when each of these combinations has a limiting proportion and within each study, the base p-values have identical distributions under the null, and identical distributions under the non-null, such as a mixture driven by random underlying effect sizes. Specifically, 
for any $\boldsymbol{c}\in\{0,1\}^n$ representing one of 
the $2^n$ combinations, let $m_{\boldsymbol{c}}$ be the number of PC hypotheses that fall into this combination. Also, let $\mathcal{H}_{0i}$ and $\mathcal{H}_{1i}$ be the sets of true nulls and true non-nulls for the $i$th study. If (a) $\lim_{M \to \infty}m_{\boldsymbol{c}} / M$ exists for all $\boldsymbol{c}$ and, (b) for each $i$, $\{P_{ij}: j \in \mathcal{H}_{0i}\}$ have identical distributions across $j$ and $\{P_{ij}: j \in \mathcal{H}_{1i}\}$ also have identical distributions across $j$,
then \Cref{assp:asymptotic} is satisfied.

\vspace{4mm}\noindent Under \Cref{assp:asymptotic}, we denote
\begin{align*}
&{\tilde{F}(\gamma)}=\pi_0 \tilde F_0(\gamma)+(1-\pi_0)\tilde F_1(\gamma),\\
&{\tilde{S}(\gamma)}=\pi_0 \tilde S_0(\gamma)+(1-\pi_0)\tilde S_1(\gamma),
\end{align*}
and further define the ``asymptotic FDR'' for a given $\gamma$ as
$${f^\infty(\gamma)} = \begin{cases}
\frac{\gamma \tilde{F}(\gamma)}{\tilde{S}(\gamma)}, \quad & \text{if }  \tilde{S}(\gamma) > 0\\
0, \quad &\text{otherwise,}
\end{cases}
$$
and the largest $\gamma_0^\infty$ such that $f^\infty(\gamma)\le\alpha$, i.e.,
$$\gamma_0^\infty = \sup\{\gamma:\ f^\infty(\gamma)\le\alpha\}.$$

Then $f^\infty(\gamma)$ is $0$ when $\gamma = 0$ and exceeds $1$ when $\gamma = 1$, thus the above set is not empty. We make a final technical assumption on the functions $f^\infty(\cdot)$, $\tilde S_0(\cdot)$ and $\tilde S_1(\cdot)$ around $\gamma_0^\infty$:

\begin{assumption}[Technical conditions]\label{assp:technical}  The following two conditions hold:
\begin{enumeratealpha}
\item There exists $\delta > 0$ such that $f^\infty(\gamma)$ is monotonically increasing in the interval $(\gamma_0^\infty-\delta, \gamma_0^\infty]$, and
\item $\tilde S_0(\gamma)$ and $\tilde S_1(\gamma)$ are both continuous at the point $\gamma_0^\infty$.
\end{enumeratealpha}  
\end{assumption}

Intuitively, (a) guarantees that the limit of the AdaFilter threshold $\gamma_{0}^{\text{BH}}$ is unique when $M \to \infty$ and (b) is satisfied if there are sufficient points (selection p-values) around $\gamma_0^\infty$ when $M$ is large.  Now we are ready to state the asymptotic FDR control of AdaFilter BH.

\begin{theorem}\label{thm:BH_asymptotic_dependence}
Under Assumptions \ref{assp:weak}-\ref{assp:technical}, the AdaFilter BH procedure of \Cref{def:two_step_MTP_BH}
 satisfies
 \begin{align*}\gamma_0^{\mathrm{BH}} &\overset{p}{\to} \gamma_0^\infty,\quad\text{and}\\
\mathrm{FDP}&\overset{p}{\to}\frac{\pi_0 \tilde S_0(\gamma_0^\infty)}{\tilde{S}(\gamma_0^\infty)} \le\alpha
\end{align*}
as $M \to \infty$. Thus, AdaFilter BH asymptotically controls FDR at the nominal level $\alpha$ for the null hypotheses $\{H_{0j}^{r/n}: j = 1, 2, \cdots, M\}$.
\end{theorem}
Notice that \Cref{assp:technical}(a) implies that $f^\infty(\gamma_0^\infty) > 0$, thereby guaranteeing $\tilde{S}(\gamma_0^\infty) > 0$.

\begin{rem} \Cref{thm:BH_asymptotic_dependence} still holds if
\Cref{assp:asymptotic} is weakened to allow $\pi_0 = 0$ while $M_0 \to \infty$ and Assumption 1 is modified to: for any fixed $\gamma$, 
$$\frac{1}{M_s^2}\sum_{j\neq j' \in \mathcal{H}_s^{r/n}}\big|\Pr(P_{ij}<\gamma,P_{ij'}<\gamma) - \Pr(P_{ij}<\gamma)\Pr(P_{ij'}<\gamma)\big|\overset{M_s \to \infty}{\longrightarrow}0$$
for both $s = 0, 1$. We can not deal with $\pi_0 = 1$ as that would lead to $\tilde S(\gamma_0^\infty) = 0$ and violates \Cref{assp:technical}(a). In \Cref{sec:simu}, we show with simulations that both simultaneous error rates can be controlled in practice even when $M_0/M = 0.99$.
\end{rem}

\subsection{Lack of complete monotonicity}\label{sec:ts_mtp_pm}

The increased power of AdaFilter can lead to an unexpected power gain when combining multiple similar studies.  
Suppose that we test the involvement of $M$ genes in a disease with two studies. 
One researcher uses BH or Bonferroni separately on the $M$ base $p$-values in each study and claims that a gene is important for the pathology if it is rejected in any of the two studies.
Another researcher runs AdaFilter with $r = 2$ on the same data while claiming that a gene is selected only when 
its nulls are false in both studies. 
The second researcher has a stricter goal, however, it is possible that  she makes more discoveries than the first.

To see how this could happen, consider the toy example in \Cref{tab:demo2}a where $M = 2$. In both studies, neither of the two hypotheses can be rejected at significance level $\alpha = 0.05$ 
when using either Bonferroni or BH on each study separately. However, both AdaFilter Bonferroni and AdaFilter BH can reject $H_{01}^{2/2}$ at the same nominal level. 
This interesting phenomenon arises from the lack of monotonicity of the number of rejections in the base p-values.  A multiple testing procedure has ``complete monotonicity'' if reducing any base $p$-values can never cause any of the decisions on the null hypotheses to switch from `reject' to `accept'. 

\begin{table}[ht]
    \begin{subtable}{0.49\textwidth}
    \caption{}
\centering
 \begin{tabular}{@{}lrr|rr@{}}
    \toprule
   & \multicolumn{2}{c}{Study} & & \\
   \cmidrule(lr){2-3}
    $j$ & 1 & 2 &  $F_j$ & $S_j$\\
    \midrule
    1 & 0.04 & 0.03 & 0.03 & 0.04\\
    2 & 0.5 & 0.9 & 0.5 & 0.9\\
    \bottomrule
  \end{tabular}
\end{subtable}
    \begin{subtable}{0.49\textwidth}
    \caption{}
\centering
 \begin{tabular}{@{}lrr|rr@{}}
    \toprule
   & \multicolumn{2}{c}{Study} & & \\
   \cmidrule(lr){2-3}
    $j$ & 1 & 2 &  $F_j$ & $S_j$\\
    \midrule
    1 & 0.04 & 0.03 & 0.03 & 0.04\\
    2 & {\color{red}0.01} & 0.9 & {\color{red}0.01} & 0.9\\
    \bottomrule
  \end{tabular}
  \end{subtable}
    \caption{(a) Toy example where AdaFilter is more efficient than testing for each study separately. Values are the p-values. (b) A counterexample to show that AdaFilter violates ``complete monotonicity''. The significance level is $\alpha = 0.05$.}
    \label{tab:demo2}
\end{table}

\begin{definition} [Complete monotonicity]\label{def:complete_monotone}
 A multiple testing procedure has complete monotonicity if each decision function $\varphi_j$ is a  non-increasing function 
in all the elements of $(p_{ij})_{n \times M}$ for $j = 1, 2, \cdots, M$.
 \end{definition}

Simes', Fisher's and Bonferroni's meta-analyses have complete monotonicity. So does the BH procedure with $n = 1$. 
Heller, Bogomolov and Benjamini \cite{heller2014deciding} call this property ``stability'' and it holds for the PC tests of \cite{heller2007}.
However, AdaFilter do not satisfy complete monotonicity: lowering one of the $p$-values for gene $j$ can change the rejection of $H_{0,j'}^{r/n}$ to acceptance for $j'\ne j$.

\Cref{tab:demo2}b shows how AdaFilter does not have complete monotonicity. Compared with \Cref{tab:demo2}a, the second hypothesis has a decreased p-value in study 1 while all other p-values are kept fixed. In \Cref{tab:demo2}a, both $\gamma_0^{\text{Bon}} = \gamma_0^{\text{BH}} = 0.05$ so the first PC hypothesis is rejected. In contrast, in \Cref{tab:demo2}b $\gamma_0^{\text{Bon}} = \gamma_0^{\text{BH}} = 0.03$ so that none of the hypotheses can be rejected though it has a smaller p-value matrix. 

This lack of complete monotonicity, which might appear undesirable, in fact is at the core of the efficiency of AdaFilter.  
A larger  $P_{ij}$ can increase  $F_j$ to reduce the multiplicity burden. When only a few hypotheses are non-null---as in a sparse genomics setting---we expect lots of large $P_{ij}$. This gives AdaFilter a substantial advantage in identifying the few non-null PC hypotheses. 
From another perspective, increased base p-values may make the signal configuration across genes more similar among studies. AdaFilter can implicitly learn such similarity and utilize it to allow more rejections.



Though lacking ``complete monotonicity'', AdaFilter retains a ``partial monotonicity'' property: reducing one of the $n$ base $p$-values for test $j$ can never change the decision from reject $H_{0,j}^{r/n}$ to accept.

\begin{definition} [Partial monotonicity]\label{def:partial_monotone}
A multiple testing procedure has partial monotonicity if for all $j \in \{1, \cdots, M\}$, 
its decision function 
$\varphi_j(p_{\cdot 1}, \dots, p_{\cdot M})$ is non-increasing in all elements of 
$(p_{1j},p_{2j},\dots,p_{nj})$. 
\end{definition}
Partial monotonicity only requires the test of hypothesis $j$ to be monotone in the $p$-values
for that same hypothesis. It allows a reduction in $p_{ij'}$ for $j'\ne j$ to reverse a rejection of $H^{r/n}_{0j}$. We have the following result:

\begin{cor} \label{cor:mono}
Both the AdaFilter Bonferroni and the AdaFilter BH procedures satisfy partial monotonicity
for all null hypotheses $H_{0j}^{r/n}$, $j = 1, 2, \dots, M$.
\end{cor}

\Cref{cor:mono} indicates that AdaFilter is reasonable in a way that reducing the base p-values of the $j$th PC hypothesis indeed strengthens the evidence of replicability for the $j$th PC hypothesis, though possibly weakening the evidence of replicability for other PC hypotheses.

\subsection{Extensions and discussion of related literature} \label{sec:discuss}
\subsubsection{Comparison with other strategies}
Two directly related methods to AdaFilter are 
\cite{bogomolov2018assessing} for $n = r= 2$ and the empirical Bayes 
approach in \cite{heller2014} for controlling the Bayes FDR, both of which are designed to test for multiple PC nulls. Both methods 
were developed to improve the efficiency of the ``direct approach'' we described.
AdaFilter is similar to the method of \cite{bogomolov2018assessing} but works for any $n$ and $r$. It provides
a frequentist approach comparable to and sometimes better than \cite{heller2014}.

The procedures of \cite{bogomolov2018assessing} use a filtering step for each study based on the p-values in the other study and a selection step that rejects hypotheses that have small enough p-values in both studies. 
To maximize the efficiency, 
the authors suggest a data-adaptive threshold. For instance, 
to control FWER, they chose two thresholds $\gamma_1$ and $\gamma_2$ to satisfy 
$$\gamma_1\times\sum_{j = 1}^M 1_{P_{2j}< \gamma_2} \approx \frac\alpha2\quad\text{and}\quad
\gamma_2\times\sum_{j = 1}^M 1_{P_{1j}< \gamma_1} \approx \frac\alpha2.$$
When $\gamma_1 \approx \gamma_2$, then 
$$\gamma_1\times\sum_{j = 1}^M 1_{\min(P_{1j}, P_{2j})< \gamma_1}
\leq \gamma_1\times\sum_{j = 1}^M \big(1_{P_{1j}< \gamma_1} + 1_{P_{2j}< \gamma_1})\approx \alpha.
$$
Thus $\gamma_0^{\text{Bon}} \approx\gamma_1 \approx\gamma_2$ and AdaFilter becomes similar to their procedure. The proposed  method only applies for $n=r=2$; this simplification makes the approach less widely applicable, despite its strong theoretical guarantees. 
In addition, for $n = r = 2$, some other methods \cite{djordjilovic2019global,djordjilovic2020optimal} have also discussed powerful multiple testing procedures controlling for FWER and in \cite{liu2020large}, the authors proposed a new procedure controlling for local FDR.

In {\tt repfdr} \cite{heller2014}, the authors tried to learn the proportion of  each of the $2^n$ (or $3^n$ for sign replicability) configurations of base hypotheses, along with the distribution of some Z-values under each 
configuration. This has cost at least $O(M2^n)$ while 
AdaFilter has cost $O(Mn\log(n))$.
There are other multiple testing procedures that aim to find consistent signals across conditions \citep{urbut2019flexible,xiang2019signal,zhao2020nonparametric}, all of which use an empirical Bayes framework as in \cite{heller2014}. Compared to these methods, AdaFilter is typically faster, guarantees simultaneous error rate control and is more robust to the dependence of p-value within each study. 

Finally, there has been much other recent literature on efficient FDR control by using some special data structure as prior knowledge \citep{lei2016adapt,li2019multiple,barber2017p,bogdan2015slope} and then adaptively determining the selection threshold. AdaFilter shares some similar adaptive filtering ideas, but works directly from an $n\times M$ matrix of $p$-values without assuming any special structure and is uniquely tailored to the special nature of the PC hypotheses.  

\subsubsection{Variable $r$ and  $n$}\label{sec:variablenr}
In many  genetic problems, the $M$ genes or DNA loci can have  varying $r_j$ or $n_j$ as they may not be present in every experiment. Then the $j$th PC null hypothesis is $H_{0j}^{r_j/n_j}$.
AdaFilter procedures still work in this scenario because \Cref{lem:frac} still holds. 
We only need to replace formulas~\eqref{eq:filterp} and~\eqref{eq:selectp} by
$$F_j = (n_j - r_j + 1)P_{(r_j - 1)j} \quad\text{and}\quad S_j = (n_j - r_j + 1)P_{(r_j)j},$$
respectively. 


\subsubsection{Requiring sign replicability}
Partial conjunctions with two-sided test statistics can reject $H^{r/n}_{0j}$ in settings where some of the significant findings have test statistics with positive signs and others negative. It is more natural to think of replication as having concordant signs, be either consistently positive or consistently negative.  In meta-analysis, one can pool $n$ one-sided tests for positive alternatives, repeat that for negative alternatives and double the smaller of the resulting one-sided $p$-values \cite{pearson-revisited}. 
This approach is very effective when either the most likely or most useful alternatives to the null have concordant signs. We can adapt this approach to PC tests and AdaFilter as follows.

We start with two base P-value matrices, $(P_{ij}^+)_{n \times M}$ and $(P_{ij}^-)_{n \times M}$, for null hypotheses $(H_{0ij}^+)_{n \times M}$ and $(H_{0ij}^-)_{n \times M}$ respectively. The rejection of $H_{0ij}^+$ is for a positive sign of the signal and  the rejection of $H_{0ij}^-$ is for a negative sign.
We also define two vectors of PC hypotheses $\{H_{01}^{r/n, +}, \dots, H_{0M}^{r/n, +}\}$ and $\{H_{01}^{r/n, -}, \dots, H_{0M}^{r/n, -}\}$. The PC null $H_{0j}^{r/n, +}$ is rejected if the signal $j$ is positive in at least $r$ studies, and $H_{0j}^{r/n, -}$ is rejected if the signal $j$ is negative in at least $r$ studies. If $r>n/2$ then it will be impossible to reject both $H^{r/n,+}_{0j}$ and $H^{r/n,-}_{0j}$ for the same $j$.


We can apply AdaFilter twice, separately on $\{H_{01}^{r/n, +}, \dots, H_{0M}^{r/n, +}\}$ and $\{H_{01}^{r/n, -},$ $
\dots, H_{0M}^{r/n, -}\}$, controlling the simultaneous error rate (FWER, PFER or FDR) at levels $\alpha_1$ and $\alpha_2$ respectively, with $\alpha_1+\alpha_2=\alpha$ (ordinarily $\alpha_1=\alpha_2=\alpha/2$). Let the set of rejected PC nulls be $\mathcal{R}^+$ and $\mathcal{R}^-$, respectively. Rejecting the union of these two sets 
$\mathcal{R}^{\pm} = \mathcal{R}^+ \cup \mathcal{R}^-$
controls the corresponding error rate at a level $\alpha = \alpha_1 + \alpha_2$ for the null hypotheses $\{H_{01}^{r/n, \pm}, \dots, H_{0M}^{r/n, \pm}\}$. 

If $r\le n/2$, then there might be some $j\in\mathcal{R}^+\cap\mathcal{R}^-$. While such findings are
not what we usually have in mind with replication they could nonetheless be scientifically interesting.


\subsubsection{Testing for all possible values of $r$}\label{sec:allr}
The partial conjunction null $H_0^{r/n}$  can be meaningfully defined whenever $2 \leq r \leq n$, and sometimes it is of interest to test for all possible $r$ values, adding another layer of multiplicity.
In \cite{benjamini2008}, it is shown that as the PC p-values $P_{j}^{r/n}$ are monotone increasing when $r$ increases, the ``direct approach'' can control for multiple $r$ values simultaneously, without any further multiplicity adjustment of $r$.  
Unfortunately, this is not true for AdaFilter. As the filtering information learnt by AdaFilter varies for different $r$ values, a signal that is rejected by a larger $r$  using AdaFilter is not guaranteed to also be rejected at a smaller replicability level. The current formulation of AdaFilter is therefore not suited to data dependent selection of the $r$ value, but requires this to be specified by the user. 

\section{Simulations}\label{sec:simu}
We benchmark the performance of AdaFilter versus the ``direct approach'' with the three forms of PC p-values in \Cref{sec:classical}. For FDR control, we also include \cite{heller2014}, using their R package {\tt repfdr}. 
Within each study, we assume a block dependence structure while changing the block size to create two scenarios, weak dependence with a small block size and strong dependence with a large block size.
 
We set $M = 10{,}000$ and consider six different configurations  
of $n$ and $r$, as listed in Table~\ref{tab:simu-table}a.
 For a given $n$, there are $2^n$ combinations of base hypotheses. In generating different configurations of the truth, we use two parameters to control the probability of each combination: $\pi_{00}$ is the probability of the global null combination and $\pi_{1}$ is the probability of the combinations not belonging to $H_{0j}^{r/n}$. We set $\pi_{1} = 0.01$ and consider two values for $\pi_{00}$: $0.8$ or $0.98$, to mimic the signal sparsity in gene expression and genetic regulation studies. 
All PC null combinations except for the global null have equal probabilities adding up to $1 - \pi_{00} - 
\pi_{1}$. All non-null PC combinations also have equal probabilities.

 


We assume that  p-values belonging to different studies are independent and,  
within one study, the correlation of the $M$ Z-values is $I_{b\times b}\otimes \Sigma_\rho $ 
where $\otimes$ is the Kronecker product.
The covariance block $\Sigma_\rho \in \real^{M/b \times M/b}$ has $1$s on the diagonal and common value $\rho = 0.5$ off 
the diagonal. We set 
the number of blocks $b = 100$ for weak dependence and $b = 10$ for strong dependence, which should cover the spectrum of what is typically expected in genomics.
When the base hypothesis is non-null, we sample the mean of its Z-value uniformly and independently from 
$\mathcal{I} = \{\pm \mu_1, \pm \mu_2, \pm \mu_3, \pm\mu_4\}$ 
where the four levels of signals $\{\mu_1, \mu_2, \mu_3, \mu_4\}$ correspond to detection power of $0.02, 0.2, 0.5, 0.95$  respectively.

In the analysis, we target controlling PFER at the nominal level $\alpha = 1$,
FDR at the nominal level $\alpha = 0.2$, and 
Bayes FDR at the same level $\alpha = 0.2$ for {\tt repfdr}. Bayes FDR  corresponds to the posterior  probability of a null hypothesis given the test statistics falling into the rejection region,  
 which has been shown to be similar to the frequentist FDR under independence \citep{efron2012}. 
 Studying PFER control, we compare four methods: AdaFilter Bonferroni and three forms of the ``direct approach''. For FDR control, we compare $6$ methods: AdaFilter BH, AdaFilter BH with the inflation factor $C(M) = \sum_{j = 1}^M 1/j \approx \log M$,  {\tt repfdr} and the ``direct approaches''.  For each parameter configuration, we run $B = 100$ random experiments and calculate the average power, number of false discoveries and false discovery proportions of each procedure. 

\Cref{tab:simu-table}b shows the average PFER and recall over the six combinations of $n$ and $r$ for each setting of $b$ and $\pi_{00}$. More detailed results for each $n$ and $r$ separately are shown in Figures S1--S2.  All methods that target PFER successfully control it at the nominal level, while the direct approaches are much more conservative, especially when both $n$ and $r$ are large. The gain in power is more pronounced when $\pi_{00}$ is higher, which is expected in many genetics applications. 

\Cref{tab:simu-table}c shows the average FDR and recall over the six combinations of $n$ and $r$ for each setting of $b$ and $\pi_{00}$. More detailed results for each $n$ and $r$ separately are shown in Figure S3--S4.  AdaFilter BH, even not inflated, and the  ``direct approach'' control FDR at the nominal level. However, similar to the  PFER control, the ``direct  approach'' procedures are too conservative. The inflated AdaFilter BH has lower power than AdaFilter BH, while its power still exceed the ``direct approach'', especially for large $r$. The {\tt repfdr} method fails to consistently control FDR especially when $n$ is large: we believe that this is due to the large number of parameters that need  to be estimated in these scenarios. In the cases when {\tt repfdr} does control FDR, its power is comparable to AdaFilter when $\pi_{00} = 0.8$ while is less when $\pi_{00} = 0.98$ is large and further reduces when dependence increases. 

Finally, we point out that our simulations only compare different methods for a pre-defined $r$ value. As discussed in Section~\ref{sec:allr}, AdaFilter needs another layer of multiplicity adjustment if multiple $r$ values are tested simultaneously.  In practice, if one aims to testing for mulitple replicability levels or is interested in obtaining the lower bound of $r$ for each hypotheses  \cite{jaljuli2019quantifying}, the ``direct approach'' may still be a preferred method as it automatically controls for the error rates of multiple $r$ values simultaneously.

\begin{table}[t]
   \centering
   \begin{subtable}{\textwidth}
   \caption{Configurations of $n$ and $r$}
   \centering
   \scalebox{0.9}{
\begin{tabular}{@{}lrrrrrr@{}}
  \toprule
  n & 2& 4 & 8 & 4 & 8 & 8\\
  r & 2& 2 & 2 & 4 & 4 & 8\\
  \bottomrule
\end{tabular}
} \end{subtable}
\newline
\vspace*{0.4 cm}
\newline
   \begin{subtable}{\textwidth}
   \centering
     \caption{Comparison of methods targeting a nominal PFER of $\alpha = 1$}
  \scalebox{0.8}{
  \begin{tabular}{@{}lcclcclcclcc@{}}
    \toprule
  & \multicolumn{5}{c}{$\pi_{00} = 0.8$} & & \multicolumn{5}{c}{$\pi_{00} = 0.98$}\\
    \cmidrule(lr){2-6}
    \cmidrule(lr){8-12}
    & \multicolumn{2}{c}{$b = 100$} & & \multicolumn{2}{c}{$b = 10$} &
    & \multicolumn{2}{c}{$b = 100$} &  & \multicolumn{2}{c}{$b = 10$}\\
    \cmidrule(lr){2-3}
    \cmidrule(lr){5-6}
 \cmidrule(lr){8-9}
    \cmidrule(lr){11-12}
    Method     & PFER & Recall($\%$)  & &PFER & Recall($\%$) & &PFER & Recall($\%$)  & &PFER & Recall($\%$)\\
    \midrule
Bon-$P_{r/n}^B$ & 0.04 & 14.72 & & 0.05 & 14.87 & & 0.00 & 14.72 & & 0.00 & 14.83 \\ 
  Bon-$P_{r/n}^F$ & 0.05 & 19.30 & & 0.06 & 19.50 & & 0.01 & 19.18 & & 0.00 & 19.38 \\ 
Bon-$P_{r/n}^S$ & 0.04 & 14.80 & & 0.05 & 14.93 & & 0.00 & 14.78 & & 0.00 & 14.88 \\ 
AdaFilter Bonferroni & 0.73 & 28.71 & & 0.76 & 28.93 & & 0.29 & 38.10 & & 0.21 & 38.25 \\ 
    \bottomrule
  \end{tabular}
  }
  \end{subtable}
  \newline
\vspace*{0.4 cm}
\newline
  \begin{subtable}{\textwidth}
  \centering
  \caption{Comparison of methods targeting a nominal FDR of $\alpha = 0.2$}
  \scalebox{0.85}{
  \begin{tabular}{@{}lcclcclcclcc@{}}
\toprule
& \multicolumn{5}{c}{$\pi_{00} = 0.8$}                             &  & \multicolumn{5}{c}{$\pi_{00} = 0.98$}                            \\ \cmidrule(lr){2-6} \cmidrule(l){8-12} 
                      & \multicolumn{2}{c}{$b = 100$} &  & \multicolumn{2}{c}{$b = 10$} &  & \multicolumn{2}{c}{$b = 100$} &  & \multicolumn{2}{c}{$b = 10$} \\ \cmidrule(lr){2-3} \cmidrule(lr){5-6} \cmidrule(lr){8-9} \cmidrule(l){11-12} 
Method                & FDR       & Recall(\%)      &  & FDR        & Recall(\%)      &  & FDR       & Recall(\%)      &  & FDR        & Recall(\%)      \\ \cmidrule(r){1-6} \cmidrule(l){8-12} 
BH-$P_{r/n}^B$        & 0.01      & 29.50           &  & 0.01       & 29.55           &  & 0.00      & 29.04           &  & 0.00       & 29.10           \\
BH-$P_{r/n}^F$        & 0.01      & 32.94           &  & 0.01       & 32.80           &  & 0.00      & 32.68           &  & 0.00       & 32.74           \\
BH-$P_{r/n}^S$        & 0.01      & 29.68           &  & 0.01       & 29.70           &  & 0.00      & 29.16           &  & 0.00       & 29.28           \\
repfdr                & 0.33      & 59.39           &  & 0.29       & 23.53           &  & 0.14      & 24.31           &  & 0.13       & 11.56           \\
AdaFilter BH          & 0.15      & 58.64           &  & 0.14       & 58.71           &  & 0.06      & 71.27           &  & 0.06       & 71.49           \\
Inflated AdaFilter BH & 0.02      & 34.39           &  & 0.01       & 34.22           &  & 0.01      & 45.70           &  & 0.01       & 46.17           \\ \bottomrule
\end{tabular}

  }
  \end{subtable}
  \caption{Simulation results. (a) lists $6$ different $n$ and $r$ scenarios considered in the simulation. (b) and (c) compare the average error rates and recalls across all $6$ $n$ and $r$ combinations under different $b$ and $\pi_{00}$ values. The results for each $n$ and $r$ are shown in Figure S1 - S4.}
\label{tab:simu-table}
\end{table}

\section{Case studies}\label{sec:real}

We apply AdaFilter to analyze two datasets: one investigates the replication of gene differential expression results in four microarray experiments
of Duchenne muscular dystrophy and one focuses on identifying marker genes of one T cell subtype from lung cancer tumors using single-cell RNA-sequencing (scRNA-seq) data. In Section S2, we also discuss the application of AdaFilter BH to a third dataset, testing for consistently significant signals 
across different metabolic super-pathways within one study.


\subsection{Duchenne Muscular Dystrophy microarray studies}

Following \cite{kotelnikova2012}, we investigate four independent Duchenne muscular dystrophy (DMD)-related microarray datasets in the Gene Expression Omnibus (GEO) database
(GDS 214, GDS 563, GDS 1956 and GDS 3027, \Cref{tab:GEO_data}a), to understand the signature genes for the disease. The goal here is to find differentially expressed marker genes for DMD that show replicating signals in multiple datasets.
For each experiment, the data is preprocessed using a standard data reprocessing tool {\tt RMA} \citep{Irizarry2003} for microarrays. Within each study, we find genes that are differentially expressed between the disease and healthy group, using a popular software {\tt Limma} \citep{smyth2004} and adjust for covariates like batch and patients' age and gender when they are available. 

The four datasets are from three different microarray platforms where different probe-sets are used. In order to compare across platforms,  we map probe-sets to common gene names. When multiple probe-sets map to the same gene, a Bonferroni rule is applied combining p-values of these probe sets into a single p-value for the gene. 
There are only $M = 1871$ genes present in all four studies, with $M = 9848$ genes shared in at least $3$ studies and $M = 13912$ genes in at least two studies. As discussed in \Cref{sec:variablenr}, AdaFilter can work with varying $n_j$ thus allow missing entries in the p-value matrix.


\begin{table}[t]

\begin{subtable}{\textwidth}
 \caption{GEO datasets information}
   \centering
  \scalebox{0.84}{
  \begin{tabular*}{0.72\columnwidth}{@{}lrrr@{}}
    \toprule
    GEO ID & Platform & Description & Source\\
    \midrule
    GDS 214 & custom Affymetrix & 4 healthy, 26 DMD & Muscle\\
    GDS 563 & Affymmetrix U95A & 11 healthy, 12 DMD & Quadriceps Muscle\\
    GDS 1956 & Affymetrix U133A & 18 healthy, 10 DMD & Muscle\\
    GDS 3027 & Affymetrix U133A & 14 healthy, 23 DMD & Quadriceps Muscle\\
    \bottomrule
  \end{tabular*}
  }
\end{subtable}
\newline
\vspace*{0.4 cm}
\newline
\begin{subtable}{0.34\textwidth}
 \caption{AdaFilter BH rejections}
\centering
\scalebox{0.9} {
 \begin{tabular}{@{}lrr@{}}
    \toprule
    $r$ & $M$ &  Rejected \\
    \midrule
    2 & 13912 &  494\\
    3 & 9848 &  142 \\
   4 & 1871 &   32 \\
    \bottomrule
  \end{tabular}
  }
\end{subtable}
\begin{subtable}{0.62\textwidth}
\caption{Known marker genes detected by AdaFilter at $r = 4$}
\centering
\scalebox{0.83} {
\begin{tabular}{@{}lrrrr@{}}
  \toprule
Gene Symbol & GDS 214 & GDS 563 & GDS 1956 & GDS 3027 \\ 
  \midrule
   \textit{MYH3} & 5.47e-14 & 2.18e-69 & 3.31e-07 & 2.49e-20 \\ 
  \textit{MYH8} & 5.74e-06 & 9.09e-11 & 2.58e-03 & 5.16e-33 \\ 
  \textit{MYL5} & 8.97e-04 & 3.06e-06 & 1.87e-03 & 6.63e-08 \\ 
  \textit{MYL4} & 1.48e-06 & 7.94e-08 & 1.21e-02 & 2.66e-08 \\  
   \bottomrule
\end{tabular}
}
\label{tab:dmd_result}
\end{subtable}
\caption{Replicability analysis for DMD microarrays}
\label{tab:GEO_data}
\end{table}

The application of AdaFilter BH at level $\alpha = 0.05$ leads to the discovery of many consistently differentially expressed genes at $r = 2, 3, 4$ (\Cref{tab:GEO_data}b). Specifically, at $r = 4$, AdaFilter BH finds $32$ significant genes (Table S2). By contrast, a BH adjustment on the Fisher combined PC p-values ($P_{r/n, j}^F$) only detects two genes (\textit{MYH3} and \textit{S100A4}) and {\tt repfdr} reports no significant genes as it fails to perform the distribution estimation of p-values with $M = 1871$ being too small. 
\Cref{tab:GEO_data}c shows four of the $32$ genes that are known to play important roles in muscle contraction (Table S1). 
Notice that besides \textit{MYH3}, all three markers do not have a small enough p-value in the third study (GDS1956, which is the least powerful study) to be detected when BH is applied to the study alone with a nominal FDR level $0.05$. However, 
AdaFilter can compensate for this deficiency by leveraging 
the overall similarity of the results in this study compared with other studies. 

\subsection{scRNA-seq of T cells in lung cancer tumors}

Understanding T cell heterogeneity in tumors brings in key information to cancer immunotherapies, and the recent single-cell RNA-sequencing (scRNA-seq) technology enables measurement of gene expression levels at the single cell resolution. In \citep{guo2018global}, the authors sequenced tumor T cells from $14$ treatment-naïve non-small-cell lung cancer patients and one main finding is the discovery of a new subtype of the CD4+ regulatory T cells (Tregs), named the suppressive tumor-resident Tregs (CD4-C9-CTLA4), that is different from the normal Tregs (CD4-C8-FOXP3). We download data from the GEO database (GSE99254), 
where cell type labels are also provided.

In order to characterize the new cell type CD4-C9-CTLA4, one need to identify a list of reliable marker genes that are consistently highly expressed in CD4-C9-CTLA4 across multiple patients. Thus we apply AdaFilter treating each patient as a ``study''.
For each patient, we obtain p-values of each gene for whether the gene expression is higher in CD4-C9-CTLA4 than in CD4-C8-FOXP3. These one-sided base p-values are calculated using the Wilcoxon rank-sum test, which is the standard test for analyzing scRNA-seq. Two patients who have less than $10$ Treg cells in either of the two groups are excluded from the analysis. In summary, we obtain a p-value matrix for $23459$ genes and $n = 12$ patients. 

\begin{figure}[ht]
\centering
\includegraphics[width = \textwidth]{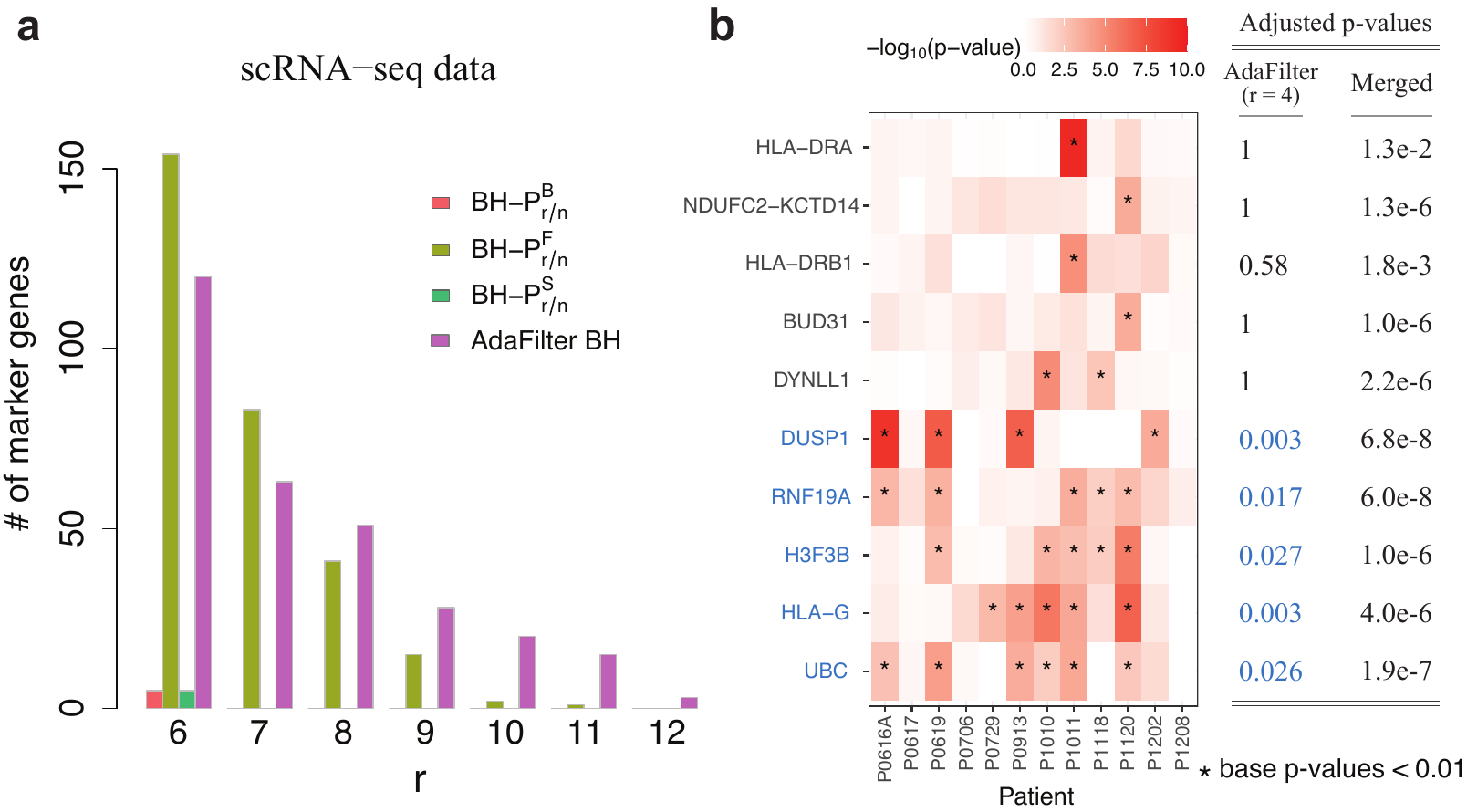}
\caption{(a) scRNA-seq data: the  number  of  genes  whose $H^{r/n}_{0j}$ were rejected by each of the compared procedures.  FDR is controlled at $\alpha = 0.05$. (b) The left is a heatmap of each patient's one-sided Wilcoxon rank-sum p-values for $10$ genes. The darker color represents a smaller p-value and a `*' label is added if it is smaller than $0.01$. The right table shows the adjusted p-values of each gene. The first column contains the adjusted AdaFilter BH p-values for $H^{4/12}$ and the second column contains the standard BH adjusted merged p-values combining cells in all patients. 
}
\label{fig:sc_comp}
\end{figure}

We vary the replicability level $r$ and \Cref{fig:sc_comp}a compares the number of genes detected using different methods. For large $r (r\geq 8)$, AdaFilter is more powerful than the ``direct approach'' with Fisher's PC p-values. However, it is less powerful when $r$ is relatively small, as the power gain of Fisher's combination to construct PC p-values may exceed the power gain using AdaFilter, whose selection p-values are from the Bonferroni's combination. The other two forms of ``direct approach'' show limited power for all $r$ and {\tt repfdr} fails to run with insufficient memory for $r\geq 6$ even with $300$G of RAM. In Table S3, we list the $20$ genes that are detected at $r = 10$, most of which are known to be linked to immunoresponse in tumors.

To further show the benefit of requiring replicability on marker gene selection, we compare a list of genes on their base p-values per patient, their standard BH adjusted merged p-values and AdaFilter BH adjusted p-values at $r = 4$ (\Cref{fig:sc_comp}b). 
All $10$ genes in \Cref{fig:sc_comp}b would be selected in the original paper as their adjusted merged p-values are far less than $0.05$. However, the top $5$ genes only have one or two patients whose base p-values are less than $0.01$. Intuitively, they are less convincing markers as there is no replicability across patients. While the merged p-values can not distinguish the more convincing markers, they can easily be separated with their AdaFilter BH adjusted p-values.

\section{Conclusion}\label{sec:conc}

Testing PC hypotheses provides a framework to detect consistently significant signals across multiple studies, leading to an explicit assessment of the replicability of scientific
findings. We introduced AdaFilter, a multiple testing procedure
which greatly increases the power in simultaneous testing of PC hypotheses over other existing methods. 
AdaFilter implicitly learns and utilizes the overall similarity of results across studies and exhibits  a lack of complete monotonicity.

We proved that AdaFilter procedures control FWER and FDR under independence of all $p$-values for a given finite number of hypotheses, and further showed that AdaFilter BH asymptotically controls FDR allowing weak dependence within each study. 
In our simulations, we demonstrated that both AdaFilter Bonferroni and AdaFilter BH are robust to the dependence of p-values within each study in practice, even when such dependence is not weak. On the other hand, 
the validity of AdaFilter does need independence of the base p-values across different studies, as Lemma~\ref{lem:frac} can be easily violated when these base p-values are dependent.


We applied AdaFilter to three case studies, encompassing gene expression and genetic association. 
Other types of applications include eQTL studies and multi-ethnic GWAS (such as new Population Architecture using Genomics and Epidemiology (PAGE) study) where it is of great interest to understand which genetic regulations are shared  and which are tissue / population specific. 
Actually, PC tests can 
be quite useful in even broader context.
According to Hume \citep{hume2003}, ``constant conjunction'' is a characteristic of causal effects. If some hypotheses are rejected repeatedly under various distinct settings, that can be supportive evidence for some causal mechanism instead simple associations. These directions can be further investigated in future research.

 \section*{Acknowledgement}
 This work was supported by the National Science Foundation under grants IIS-1837931, DMS-1521145 and DMS-2113646, and by the National Institutes of Health under grants R01MH101782 and U01HG007419.
 We thank Y. Benjamini, M. Bogomolov, R. Heller and N. R. Zhang for helpful discussions.

\begin{supplement}
\textbf{Supplementary information (SI)}.
The SI includes supplementary text (Sections S1-S3), Table S1-S4 and Figure S1-S6.
\end{supplement}

\bibliographystyle{imsart-number}
\bibliography{ref.bib,moreref.bib}

\newpage

\renewcommand{\thesection}{S\arabic{section}}   
\renewcommand{\thetable}{S\arabic{table}}   
\renewcommand{\thefigure}{S\arabic{figure}}

\setcounter{section}{0}
\setcounter{figure}{0}
\setcounter{table}{0}

\renewcommand {\thepage} {S\arabic{page}}
\setcounter{page}{1}

\title{Supplementary Information for  \\``Detecting Multiple Replicating Signals using Adaptive Filtering Procedures''}

\begin{center}
    Jingshu Wang, Lin Gui, Weijie J. Su, Chiara Sabatti, Art B. Owen
\end{center}

\section{Combining AdaFilter Bonferroni with the sequential rejection principle}

As a procedure to control FWER, we can apply the sequential rejection principle \citep{goeman2010sequential} on AdaFilter Bonferroni to further increase the number of rejections while controlling FWER at the same level. As discussed in \cite{goeman2021only}, an alternative approach to improve the power of AdaFilter Bonferroni is to apply the closed testing procedure. We will introduce these two different approaches for AdaFilter Bonferroni and show that they are equivalent. We will also compare the power of this improvement with AdaFilter Bonferroni via simulations.  

\subsection{Sequential AdaFilter Bonferroni}
Let $\mathcal{R}_k\subseteq \{1, 2,\dots,M\}$, $k=0,1,\dots$, be the rejection set after step $k$. The sequential AdaFilter Bonferroni is defined as follows:
\begin{equation}
\label{Eq:sequential}
\begin{split}
\mathcal{R}_0=&\varnothing\\
\mathcal{R}_{k+1}=&\mathcal{R}_{k}\cup\mathcal{N}(\mathcal{R}_k),
\end{split}
\end{equation}
where $\mathcal{N}:2^{\{1, \cdots, M\}}\to\{1, \cdots, M\}$ is the successor function defined by
\begin{equation}
\label{Eq:sequential}
\begin{split}
\gamma_0(\mathcal{R}) =& \sup\{\gamma\in[0,\alpha]:\gamma\sum_{j\in\mathcal{R}^c}1_{F_j<\gamma}\le\alpha\}\\
\mathcal{N}(\mathcal{R})=&\{j\in\mathcal{R}^c:S_j<\gamma_0(\mathcal{R})\},
\end{split}
\end{equation}
The final rejection set is given by $\mathcal{R}_\infty=\lim_{k\to\infty}\mathcal{R}_k$. Intuitively, after removing the rejected PC hypotheses from AdaFilter Bonferroni, we can apply AdaFilter Bonferroni again on the remaining PC hypotheses at the same significance level $\alpha$ to reject more hypotheses, and continue this process until we can not reject any more hypotheses. The final rejection set of this sequential adaFilter Bonferroni is the union of the rejection hypotheses at all steps. 

In order to prove that this sequential adaFilter Bonferroni controls FWER at level $\alpha$, we make use of Theorem 1 in \cite{goeman2010sequential}, which guarantees an $\alpha$-level FWER control of a sequential rejection procedure under the following two conditions:
\begin{enumerate}
\item (monotonicity condition) For every fixed $\mathcal{R}\subseteq\mathcal{S}\subset\{1, 2, \cdots, M\}$
$$\mathcal{N}(\mathcal{R})\subseteq\mathcal{N}(\mathcal{S})\cup\mathcal{S}$$
\item (single-step condition) Denote $\mathcal{H}_1^{r/n} = \{j: H_{0j}^{r/n} \text{is false}\}$ as the set of hypotheses whose PC null is false, then $$P(\mathcal{N}(\mathcal{H}_1^{r/n})\subseteq\mathcal{H}_1^{r/n})\ge1-\alpha.$$ 
\end{enumerate}
We show that these two conditions are satisfied for the sequential AdaFilter Bonferroni.

First, we show that the monotonicity condition holds. For every $\mathcal{R}\subseteq\mathcal{S}\subset \{1, \cdots, M\}$, we have $\mathcal{R}^c\supseteq\mathcal{S}^c$. So, for any fixed $\gamma\in[0,\alpha]$, if $\gamma\sum_{j\in\mathcal{R}^c}1_{F_j<\gamma}\le\alpha$, we must have
$$\alpha\ge\gamma\sum_{j\in\mathcal{R}^c}1_{F_i<\gamma}\ge\gamma\sum_{j\in\mathcal{S}^c}1_{F_j<\gamma},$$
which means $\left\{\gamma\in[0,\alpha]:\gamma\sum_{j\in\mathcal{R}^c}1_{F_j<\gamma}\right\}\subseteq\left\{\gamma\in[0,\alpha]:\gamma\sum_{j\in\mathcal{S}^c}1_{F_j<\gamma}\right\}$ and hence $\gamma_0(\mathcal{S})\ge\gamma_0(\mathcal{R})$. Since $\forall j\in\mathcal{N}(\mathcal{R})$, $j$ is either in $\mathcal{S}$ or $j$ satisfies with the condition that $j\in\mathcal{S}^c$ and $S_j<\gamma_0(\mathcal{S})$, which is equivalent to $j\in\mathcal{N}(\mathcal{S})$. Thus, we get $$\mathcal{N}(\mathcal{R})\subseteq\mathcal{N}(\mathcal{S})\cup\mathcal{S}.$$ 

Then we show that the single-step condition holds. Notice that 
$$P(\mathcal{N}(\mathcal{H}_1^{r/n})\subseteq\mathcal{H}_1^{r/n})\ge1-\alpha\Leftrightarrow P(\cup_{i\in\mathcal{H}_0^{r/n}}\{S_i<\gamma_0(\mathcal{H}_1^{r/n})\})\le\alpha$$ 
where $\mathcal{H}_0^{r/n} = \{j: H_{0j}^{r/n} \text{is true}\}$. 
Since for every set $\mathcal{R}$, Theorem 4.2 guarantees 
$$P(\cup_{j\in\mathcal{R}^c\cap\mathcal{H}_0^{r/n}}\{S_j<\gamma_0(\mathcal{R})\})\le\alpha.$$
We set $\mathcal{R}=\mathcal{H}_0^{r/n}$, then we get
$$P(\cup_{j\in\mathcal{H}_0^{r/n}}\{S_j<\gamma_0(\mathcal{H}_1^{r/n})\})\le\alpha,$$
and the single-step condition is guaranteed.

\subsection{Closed AdaFilter Bonferroni}
An alternative approach to improve the power of AdaFilter Bonferroni is to apply the closed testing procedure \citep{goeman2021only}. To derive the closed testing procedure, for any set of hypotheses $\mathcal{S} \in \{1, \cdots, M\}$, we first define the corresponding local testing of AdaFilter Bonferroni testing for the global null of a set of hypotheses  $\mathcal{S}$ 
$$H_\mathcal{S}: H_{0j}^{r/n}\text{ true, for all }j\in\mathcal{S}.$$
A valid rejection rule for $H_\mathcal{S}$ based on AdaFilter Bonferroni is 
$$\Psi_\mathcal{S}=1_{\min_{j\in\mathcal{S}} S_j<\gamma_{0, \mathcal{S}}},$$
where $\gamma_{0, \mathcal{S}} = \sup\left\{\gamma:\gamma\sum_{j\in\mathcal{S}}1_{F_j<\gamma}\le\alpha\right\}$ (notice that it is different from $\gamma_{0}(\mathcal{S})$ defined in \eqref{Eq:sequential}). The global null $H_\mathcal{S}$ is rejected only when AdaFilter Bonferroni on $\mathcal{S}$ rejects at least one hypothesis. 

Then the closed testing procedure for AdaFilter Bonferroni is that for PC hypothesis $j$, define  
$$\phi_j=1_{\{\text{reject } H_{0j}^{r/n}\}}=\min_{\mathcal{S}: j\in\mathcal{S} \subseteq \{1, \cdots, M\}}\Psi_\mathcal{S},$$
and we reject all PC hypotheses with $\phi_j = 1$.

We now derive a more explicit description of this closed AdaFilter Bonferroni procedure. First, notice that for any two sets $\mathcal S_1$ and $\mathcal S_2$, if $\mathcal S_1 \subseteq \mathcal S_2$, then $\gamma_{0, \mathcal{S}_1} \geq \gamma_{0, \mathcal{S}_2}$. If we further have $\min_{j \in \mathcal{S}_1}S_j = \min_{j \in \mathcal{S}_2}S_j$, then $\Psi_{\mathcal{S}_1} \geq \Psi_{\mathcal{S}_2}$. Thus, for PC hypothesis $j$, we have 
$$\phi_j = \min_{j': S_{j'} \leq S_j} \Psi_{\{l: S_l \geq S_{j'}\}}.$$
Order the selection p-values as $S_{(1)}\le S_{(2)}\le\cdots\le S_{(M)}$ and denote $H_{0(k)}^{r/n}$ as the corresponding PC null of $S_{(k)}$. Also, denote $\mathcal{I}_{(k)}^+ = \{j: S_j \geq S_{(k)}\}$, then 
 $$H_{0(k)}^{r/n}\text{ is rejected}\Longleftrightarrow S_{(1)}\le\gamma_{0, \mathcal{I}^+_{(1)}},\ S_{(2)}\le\gamma_{0, \mathcal{I}^+_{(2)}},\dots,\ S_{(k)}\le\gamma_{0, \mathcal{I}^+_{(k)}}.$$
 In other words, define
$$\widehat{k}=\max\left\{k:S_{(1)}\le\gamma_{0, \mathcal{I}^+_{(1)}},\ S_{(2)}\le\gamma_{0, \mathcal{I}^+_{(2)}},\dots,\ S_{(k)}\le\gamma_{0, \mathcal{I}^+_{(k)}}\right\},$$
then the closed AdaFilter Bonferroni is to reject $H_{0(1)}^{r/n},\dots,H_{0(\widehat{k})}^{r/n}$.

In \citep{goeman2021only}, the authors proved that the closed testing procedure controlling FWER is admissible only when the local test for the global null $H_\mathcal{S}$ is admissible. For the partial conjunction problem, each $H_{0j}^{r/n}$ is a composite null and the admissibility of tests for $H_{0j}^{r/n}$ has been discussed in \cite{wang2018admissibility}. However, whether the local test based on AdaFilter Bonferroni is admissible needs further investigation.

\subsection{Equivalence of the two approaches}

We prove that the sequential AdaFilter Bonferroni and the closed AdaFilter Bonferroni are actually equivalent. 

First, notice that for any hypothesis $H_{0(k)}^{r/n}$ that is rejected by the sequential AdaFilter Bonferroni procedure, it satisfies $S_{(k)}\leq \gamma_{0,\mathcal{I}^+_{(k')}}$ for some $k'\le k$. As $\mathcal{I}^+_{(k')} \subseteq\mathcal{I}^+_{(k)}$, we have $\gamma_{0,\mathcal{I}^+_{(k')}} \leq \gamma_{0,\mathcal{I}^+_{(k)}}$. Thus, any hypotheses that is rejected by closed AdaFilter Bonferroni must be rejected by the sequential AdaFilter Bonferroni. 


On the other hand, if there are hypotheses that are rejected by closed AdaFilter Bonferroni, but not by sequential AdaFilter Bonferroni, then denote $S_{k}$ as the smallest rejection p-values of those hypotheses. All $H_{0(k')}^{r/n}$ are rejected by both approaches if $k' < k$. As it is rejected by the closed AdaFilter Bonferroni, it satisfies $S_{(k)} \leq \gamma_{0, \mathcal{I}^+_{(k)}}$. However, if that's true, from the definition of the sequential AdaFilter Bonferroni, it should also be rejected by the sequential procedure, which is a contradiction. Thus, both approaches must reject the same set of hypotheses. 


\subsection{Power comparison between the sequential and the original AdaFilter Bonferroni}
Finally, we study the power again of the sequential AdaFilter Bonferroni over the original one via simulations. The simulation setting is the same as \Cref{sec:simu} and we control FWER at level $\alpha = 0.05$. As shown in Table~\ref{tab:seq-simu}, we do observe an increase of the power in finding replicable signals after using the sequential AdaFilter Bonferroni. 

\begin{table}[ht]
 \caption{Comparison of methods targeting a nominal FWER of $\alpha = 0.05$}
 \centering
 \scalebox{0.82}{
 \begin{tabular}{@{}lcclcclcclcc@{}}
    \toprule
  & \multicolumn{5}{c}{$\pi_{00} = 0.8$} & & \multicolumn{5}{c}{$\pi_{00} = 0.98$}\\
    \cmidrule(lr){2-6}
    \cmidrule(lr){8-12}
    & \multicolumn{2}{c}{$b = 10$} & & \multicolumn{2}{c}{$b = 100$} &
    & \multicolumn{2}{c}{$b = 10$} &  & \multicolumn{2}{c}{$b = 100$}\\
    \cmidrule(lr){2-3}
    \cmidrule(lr){5-6}
 \cmidrule(lr){8-9}
    \cmidrule(lr){11-12}
    Method     & FWER & Recall($\%$)  & &FWER & Recall($\%$) & &FWER & Recall($\%$)  & &FWER & Recall($\%$)\\
    \midrule
Bon-$P_{r/n}^B$ & 0.00 & 15.06 & & 0.00	& 15.11 & & 0.00 & 15.07 &  & 0.00	& 15.13 \\ 
Bon-$P_{r/n}^F$ & 0.00 & 21.70 & & 0.00 & 21.74 & & 0.00 & 21.57 & & 0.00	& 21.66 \\ 
Bon-$P_{r/n}^S$ & 0.00 & 15.12 & & 0.00 & 15.16 & & 0.00 & 15.11 & & 0.00 & 15.17 \\ 
AdaFilter Bonferroni &  0.03 & 26.85 & & 0.03 & 26.78 & & 0.01 & 34.89 & & 0.01 & 35.09 \\ 
Sequential AdaFilter Bonferroni & 0.03 & 27.24 & &  0.04 & 27.19 & & 0.02 & 37.34 & & 0.01 & 37.71\\ 
    \bottomrule
  \end{tabular}
  }
 \label{tab:seq-simu}
\end{table}

\newpage

\section{Application of AdaFilter to the metabolites super-pathways GWAS data}

The  multi-trait GWAS data from \cite{shin2014} is a comprehensive study of  the 
genetic loci influencing  human metabolism: in addition to DNA variation, it
 measured the levels of  $333$  metabolites,  categorized into $8$ non-overlapping ``super pathways'', and integrated this data with gene expression and other prior information.
Shin et al.\ \cite{shin2014}. 
 strongly emphasize how distinct metabolic traits are linked through the effects of specific genes and indicates that the discovery of genes that affect a diverse class of metabolic measurements is particularly interesting as these genes are associated with complex trait/disease or drug responses.

Testing for partial conjunction is a means to discover such genes. Specifically, we apply AdaFilter to the  tests  for association between single nucleotide polymorphisms (SNPs) and ``super-pathways'' (each SNP is linked to a gene, and hence discovering a SNP points to a specific gene; super-pathways are defined in \cite{shin2014}).

In \cite{shin2014}, a total of $7824$ adult individuals from two European populations were recruited in the study, and  $M = 2{,}182{,}555$ SNPs were recorded, either directly genotyped or imputed from the HapMap 2 panel. 
Out of the $333$ annotated metabolite traits reported in the paper, only $275$ have the summary statistics (t statistics and p-values for the association of each SNP and trait) publicly available at the Metabolomics GWAS Server \url{http://mips.helmholtz-muenchen.de/proj/GWAS/gwas/index.php?task=download}, \\ which is the data we use for analysis.

\subsection{Calculating the p-values for each individual super-pathway}

To calculate base p-values $p_{ij}$ 
for each marker $j$ and each super-pathway $i$, we start with the  Z-values 
$Z_{sj}$ for test of association between each metabolite $s$ and marker $j$, which are given as summary statistics. For a super-pathway $i$, 
let $\{s_1, s_2, \cdots, s_{n_i}\}$ be the index set of metabolite measures that belong to it. 
We assume that $(Z_{s_1j}, Z_{s_2j}, \cdots, Z_{s_{n_i}j}) \sim \mathcal{N}(0, \Sigma_i)$. 
The covariance $\Sigma_i$ can be accurately estimated in principle  
since we have millions of markers. Most of the base hypotheses are null and the noise of the estimates of the marginal effects of these SNPs should share a common correlation matrix \citep{bulik2015atlas}. 
We estimate  
$\Sigma_i$ using graphical Lasso, assuming that the precision matrix is sparse. To do this, we  randomly sample $2000$ SNPs (markers)  that lie at least $1$Mbp away from 
each other, so that they can be considered as independent SNPs. Then these SNPs are treated as samples in Graphical Lasso and the tuning parameters of the final estimates selected by cross-validation. The Graphical Lasso approach guarantees an accurate sparse inverse covariance matrix estimation, that is needed for computing the p-values for each super-pathway. 

\begin{figure}[ht]
\includegraphics[width = 0.9\linewidth]{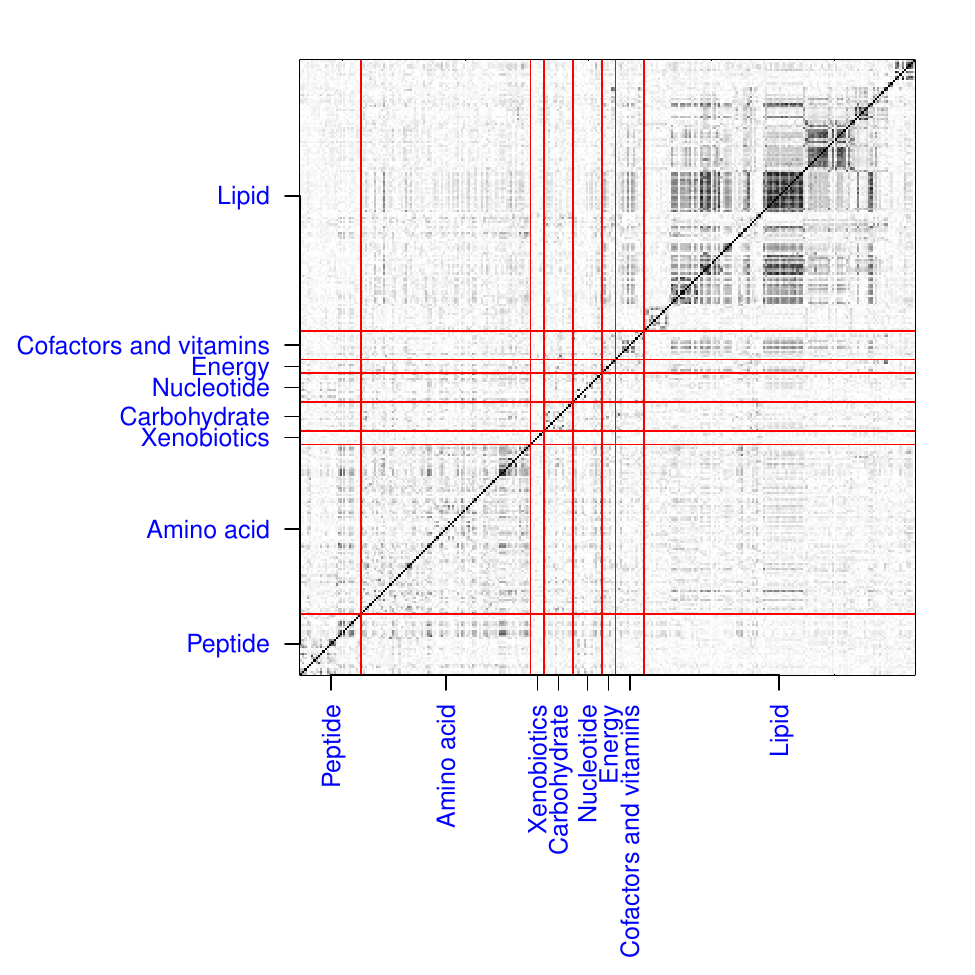}
\caption{Correlation of test statistics of the $m = 275$ metabolites. The darker the color, the higher the absolute value of the correlation. The metabolite measurements are reordered so that metabolites in the same pathway are adjacent to each other. The red lines and blue texts label the 
eight super-pathways. } 
\label{fig:cor_meta}
\end{figure}

Let $p_{ij}$ be the p-value for the association between a super-pathway $i$ and marker $j$, in other words, the 
p-value for the null 
$$H_{ij0}: \text{ no metabolite measure of super pathway } i \text{ is associated with marker } j.$$ 
Given $(Z_{s_1j}, Z_{s_2j}, \cdots, Z_{s_{n_i}j})$ and $\hat\Sigma_i$,  
we calculate p-values $p_{ij}$ from the chi-square test treating the estimated 
$\Sigma_i$ as known. These $p_{ij}$ serve as base p-values which will be used in the partial conjunction testing.
\Cref{fig:cor_meta} shows the estimated correlation across metabolites assuming $(Z_{1j}, Z_{2j}, \cdots, Z_{mj}) \sim\mathcal{N}(0, \Sigma)$ for $j = 1, 2, \cdots, M$. We estimate $\Sigma$ by applying the Minimum Covariance Determinant (MCD) \citep{rousseeuw1999} estimator to the $2000$ randomly sampled SNPs, where MCD is a highly robust method to reduce the influence of the sparse non-null hypotheses. Notice that we choose MCD instead of graphical Lasso here as we do not need an estimate of the inverse of $\Sigma$. It is evident that most of the nonzero correlations are between traits within the same super-pathways: this allows us
to apply the adaptive filtering procedures for PC hypotheses across super-pathways with confidence. 

\subsection{AdaFilter results}

\begin{figure}[ht]
\centering
\includegraphics[width = \textwidth]{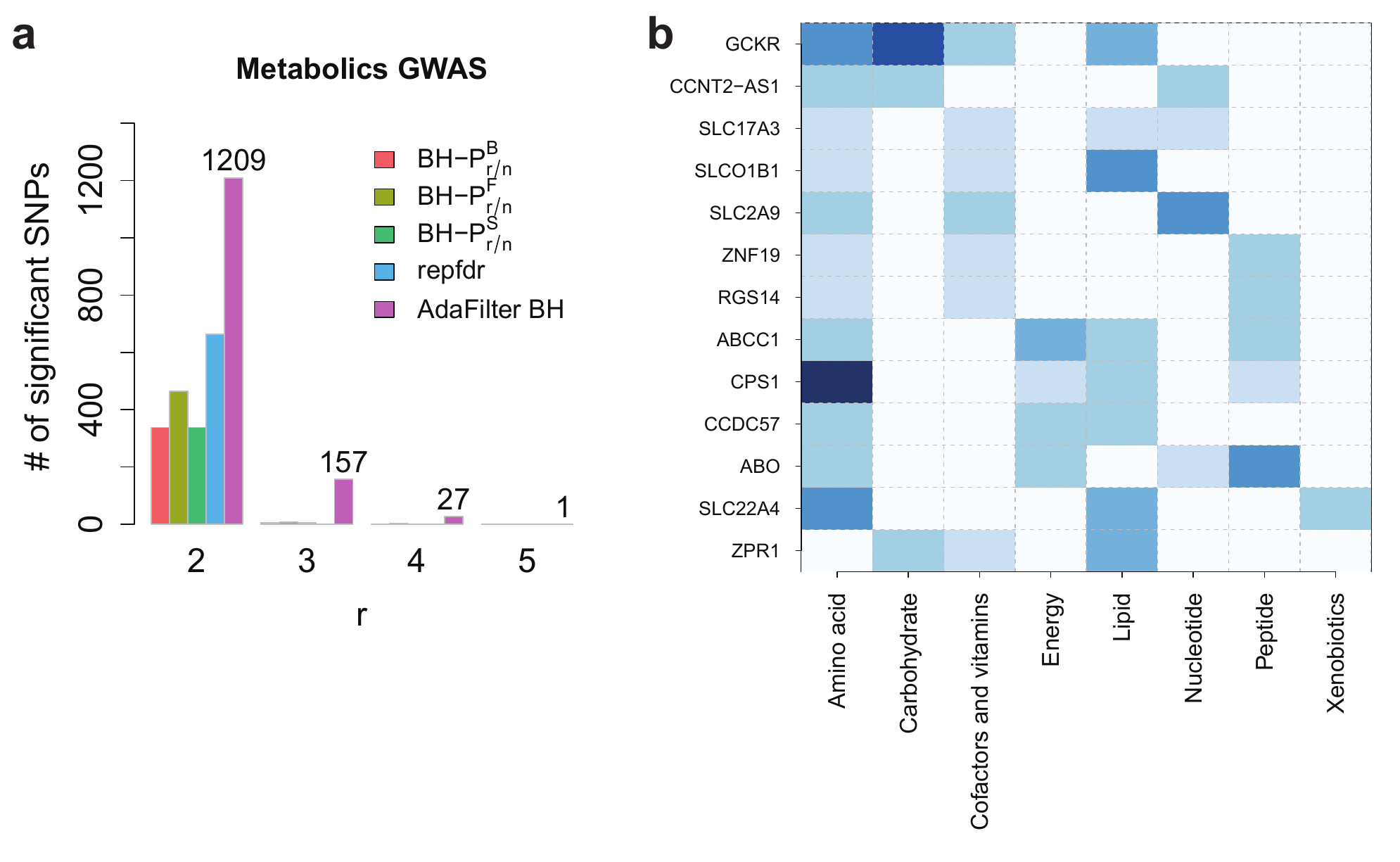}
\caption{ 
(a) Metabolics GWAS data: the number of SNPs whose $H^{r/n}_{0j}$ were rejected by each of the compared procedures.   FDR is controlled at $\alpha = 0.05$. 
(b) Visualization of base p-values of the super-pathways for the $13$ clumped significant SNPs (their mapped genes are labeled) detected at $r = 3$ (Web Table 2). For the two SNPs that map to the same gene, only the more significant one is shown. The significant p-values have a blue color. The darker the color, the smaller the p-value is.}
\label{fig:gwas}
\end{figure}

\Cref{fig:gwas}a compares the number of significant SNPs when FDR is controlled at $0.05$ and $r$ ranges from $2$ to $5$. 
Compared with other four methods, our AdaFilter BH is much more powerful for any value of $r$. 
The method {\tt repfdr} rejects less than AdaFilter BH, which is a consistent result with the simulations that {\tt repfdr} may suffer from a power deficiency under dependence structures.  

Among the significant SNPs at $r = 3$, $14$ different SNPs are detected after clumping using PLINK 1.9 (\cite{purcell2007plink}, Table S4), representing $13$ different genes (\Cref{fig:gwas}b). Many of these genes have important roles in complex disease. For instance, gene \textit{GCKR} encodes a regulatory protein that inhibits glucokinase, which regulates carbohydrate metabolism, converting glucose to amino acid and fatty acids. It is also a potential drug target for diabetes. 
Several genes (\textit{SLC17A3}, \textit{SLC2A9}, \textit{SLC22A4}, \textit{SLCO1B1}) encoding the solute carrier (SLC) group of membrane transport proteins are also 
detected.  This suggests that they might function to transport multiple solutes and could possibly be drug targets for diabetes, chronic kidney disease and various autoimmune diseases.

\newpage

\newpage
\section{Proofs}
In this section, we provide proofs for all the theoretical results in Section 3 and Section 4. In addition to the notations in the main text, 
we define  $p_{\cdot j} = (p_{1j}, \ldots, p_{nj})$ be the vector of p-values for base hypotheses $(H_{01j},\ldots,H_{0nj})$ involved in $H_{0j}^{r/n}$ for each $j = 1, 2, \ldots, M$. Also, we use $1{:}M$ as a concise notation of the index set $\{1, 2, \ldots, M\}$. For a set $u \subset 1{:}n$, define $\bm P_{ui} = \{P_{ki}: k \in u\}$ and $\bm P_{ui} \preceq \lambda$ be the event that all elements $P_{ki} \in \bm P_{ui}$ satisfy $P_{ki}\leq \lambda$ for some scalar $\lambda$. 


\subsection{Proof of \Cref{prop:equiv}} 

We first show that the set of rejections defined as $\{j: P_j^{\text{Bon}} < \alpha\}$ is the same as the rejections obtain from \Cref{def:two_step_MTP_bonf}. For any $S_{j_1} \leq S_{j_2}$, as we always have $m_{j_1}^{\text{AF}} \leq m_{j_2}^{\text{AF}}$ by definition, we also have $P_{j_1}^{\text{Bon}} \leq P_{j_2}^{\text{Bon}}$. Thus, equivalently we need to show that for any $j$
$$S_{(j)}<\gamma_0^{\text{Bon}} \Leftrightarrow P_{(j)}^{\text{Bon}} <\alpha.$$
If $S_{(j)}<\gamma_0^{\text{Bon}}$, then
$$S_{(j)}\sum_{h=1}^M1_{F_{h} \leq S_{(j)}}< \gamma_0^{\text{Bon}}\sum_{h=1}^M1_{F_{h} <  \gamma_0^{\text{Bon}}}\le\alpha,$$
thus we have $P_{(j)}^{\text{Bon}} = S_{(j)}\sum_h 1_{F_h \leq S_{(j)}}<\alpha$. On the other hand, if $P_{(j)}^{\text{Bon}}<\alpha$, by the definition of the adjusted p-values we have 
$$S_{(j)}\sum_{h=1}^M1_{F_{h}<S_{(j)}}\le S_{(j)}\sum_{h=1}^M1_{F_{h}\le S_{(j)}}<\alpha.$$
Thus, based on the definition of $\gamma_0^{\text{Bon}}$, we have $S_{(j)}\le \gamma_0^{\text{Bon}}$. If $S_{(j)}=\gamma_0^{\text{Bon}}$, then as $P_{(j)}^{\text{Bon}}<\alpha$, we also obtain
$$\gamma_0^{\text{Bon}}\sum_{h=1}^M1_{F_{h}<\gamma_0^{\text{Bon}}}<\alpha.$$
Based on the definition of $\gamma_0^{\text{Bon}}$,
$$
\alpha \le\inf_{\gamma > \gamma_0^{\text{Bon}}} \gamma\sum_{h=1}^M1_{F_{h}<\gamma} =  \gamma_0^{\text{Bon}}\sum_{h=1}^M1_{F_{h}\le\gamma_0^{\text{Bon}}}=S_{(j)}\sum_{h=1}^M1_{F_{h}\le S_{(j)}},$$
which is contradictory to the fact that $P_{(j)}^{\text{Bon}}<\alpha$. Hence, we get $S_{(j)}<\gamma_0^{\text{Bon}}$.

Then we show that the set of rejections defined as $\{j: P_j^{\text{BH}} < \alpha\}$ is the same as the rejections obtain from \Cref{def:two_step_MTP_BH}. Again, for any $S_{j_1} \leq S_{j_2}$, we also have $P_{j_1}^{\text{BH}} \leq P_{j_2}^{\text{BH}}$. Thus, equivalently we need to show that for any $j$
$$S_{(j)}<\gamma_0^{\text{BH}} \Leftrightarrow P_{(j)}^{\text{BH}}<\alpha.$$
Define $k_0^{\text{BH}}=\max\{k: S_{(k)}<\gamma_0^{\text{BH}}\}$. By definition, we have
\begin{equation}
 S_{(j)}<\gamma_0^{\text{BH}}\Leftrightarrow S_{(j)}\le S_{(k_0^{\text{BH}})}. 
 \label{Eq3}
\end{equation}
Since
$$P_{(k_0^{\text{BH}})}^{\text{BH}}=\frac{S_{(k_0^{\text{BH}})}\sum_{h=1}^M1_{F_h\le S_{(k_0^{\text{BH}})}}}{k_0^{\text{BH}}}<\frac{\gamma_0^{\text{BH}}\sum_{h=1}^M1_{F_h< \gamma_0^{\text{BH}}}}{k_0^{\text{BH}}}=\frac{\gamma_0^{\text{BH}}\sum_{h=1}^M1_{F_h< \gamma_0^{\text{BH}}}}{\sum_{h=1}^M1_{S_h<\gamma_0^{\text{BH}}}}\le\alpha,$$
we obtain
\begin{equation}
S_{(j)}\le S_{(k_0^{\text{BH}})}\Leftrightarrow P_{(j)}^{\text{BH}}\le P_{(k_0^{\text{BH}})}^{\text{BH}}<\alpha.
\label{Eq4}
\end{equation}
Additionally, if there exists some $j$ satisfying both $P_{(j)}^{\text{BH}}<\alpha$ and $P_{(j)}^{\text{BH}}> P_{(k_0^{\text{BH}})}^{\text{BH}}$, then we have $S_{(j)}\ge\gamma_0^{\text{BH}}$ and $\exists$ $k\ge j$ such that
$$\alpha>P_{(j)}^{\text{BH}}=\frac{S_{(k)}\sum_{h=1}^M1_{F_h\le S_{(k)}}}{k}.$$
Then, there exists a $\tilde \gamma$ in a small neighbourhood of $S_{(k)}$ with $\tilde \gamma>S_{(k)}\ge\gamma_0^{\text{BH}}$ such that
$$\alpha>\frac{\tilde \gamma\sum_{h=1}^M1_{F_h< \tilde\gamma}}{k}\ge\frac{\tilde\gamma\sum_{h=1}^M1_{F_h< \tilde\gamma}}{\sum_{h=1}^M1_{S_h< \tilde\gamma}},$$
which contradicts the definition of $\gamma_0^{\text{BH}}$. Thus we get 
\begin{equation}
\label{Eq5}
P_{(j)}<\alpha\Leftrightarrow P_{(j)}\le P_{(k_0^{\text{BH}})}  
\end{equation}
Combining \eqref{Eq3} - \eqref{Eq5}, we obtain 
$$S_{(j)}<\gamma_0^{\text{BH}}\Leftrightarrow P_{(j)}<\alpha. $$

\subsection{Proof of \Cref{lem:frac} (Conditional validity)}

We use $u \subseteq\{1, 2, \cdots, n\}$ to represent a subset of the studies.
This set $u$ has cardinality $|u|$. We use $-u$ to denote its
complement $\{1, 2, \cdots, n\}\backslash u$.

Equivalent to the lemma, we need to show that under $H_{0j}^{r/n}$ for any $\gamma > 0$,
$$\Pr(P_{(r)j} < \gamma) \leq (n - r + 1)\Pr(P_{(r-1)j} < \gamma)$$

By independence of $P_{ij}$ across $i$, we have the decomposition
\begin{align*}
\Pr(P_{(r)j} < \gamma) & = \sum_{k = r}^{n} \sum_{| u | = k} \prod_{i \in u}
   \Pr(P_{i j} <\gamma) \prod_{i \in - u} \Pr(P_{i j} \ge \gamma),\quad\text{and} \\
 \Pr(P_{(r - 1) j} < \gamma) &= \sum_{k = r - 1}^{n} \sum_{| u | = k} \prod_{i
   \in u} \Pr(P_{i j} < \gamma) \prod_{i \in - u} \Pr(P_{ij} \ge \gamma) .
\end{align*}

One critical observation is that, when $H_{0j}^{r/n}$ is true, for any $u\subset1{:}n$ with $|u|\geq r$
there is at least one index $i^*_u \in u$
for which $H_{0{i_u^*}j}$ is true. 
Then, we have
\begin{align*}
     \Pr(P_{(r) j} < \gamma) 
  & =  \sum_{k = r}^{n} \sum_{| u | = k} \Big(\prod_{i \in u} \Pr(P_{i j}
  < \gamma)  \prod_{i \in - u} \Pr(P_{i j} \ge \gamma)\Big)\\
  & \leq \gamma \cdot \sum_{k = r}^{n} \sum_{| u | = k}  \Big (\prod_{i \in u \setminus  \{ i^*_u \}} \Pr
  (P_{i j} < \gamma)   
  \prod_{i \in - u} \Pr(P_{i j} \ge \gamma) \Big)
  \end{align*}
To make a connection with $P_{(r-1)j}$, we make the following decomposition
\begin{align*}
   & \prod_{i \in u \setminus  \{ i^*_u \}} \Pr
  (P_{i j} < \gamma)   
  \prod_{i \in - u} \Pr(P_{i j} \ge \gamma)\\
  =& \prod_{i \in u \setminus  \{ i^*_u \}} \Pr
  (P_{i j} < \gamma)   
  \prod_{i \in - u} \Pr(P_{i j} \ge \gamma)
  \big[\Pr(P_{i^*_u j} < \gamma) + \Pr(P_{i^*_u j} \ge \gamma)\big]\\
  = & \prod_{i \in u} \Pr(P_{i j}
  < \gamma)  \prod_{i \in - u} \Pr(P_{i j} \ge \gamma)
    + \prod_{i \in u\setminus {\{i^*_u\}}} \Pr(P_{i j}
  < \gamma)  \prod_{i \in - u\cup\{i^*_u\}} \Pr(P_{i j} \ge \gamma)
  \end{align*}
Notice that for the second term we can reorganize and get 
\begin{align*}
  &  \sum_{k = r}^{n} \sum_{| u | = k} \Big(\prod_{i \in u\setminus {\{i^*_u\}}} \Pr(P_{i j}
  < \gamma)  \prod_{i \in - u\cup\{i^*_u\}} \Pr(P_{i j} \ge \gamma)\Big)\\
  \leq & \sum_{k = r - 1}^{n - 1} (n - k) \sum_{| u' | = k} \Big(
  \prod_{i \in u'} \Pr(P_{ij} < \gamma) \prod_{i \in - u'} \Pr(P_{ij} \ge \gamma)\Big)\\
\end{align*}
where the inflation $n-k$ for each $| u' | = k$ is due to the fact that there are at most $(n-k)$ different $u$ whose $|u|=k + 1$ reduce to $u'$ after removing the index $i^*_u$.
Thus combining the above results, we have 
\begin{align*}
\Pr(P_{(r)j} < \gamma) \leq   &  \gamma \cdot \sum_{k = r}^{n} \sum_{| u | = k} \Big(\prod_{i \in u} \Pr(P_{i j}
  < \gamma)  \prod_{i \in - u} \Pr(P_{i j} \ge \gamma)\Big) \\
  &  + \ \gamma \cdot \sum_{k = r - 1}^{n - 1} (n - k) \sum_{| u | = k} \Big(
  \prod_{i \in u} \Pr(P_{ij} < \gamma) \prod_{i \in - u} \Pr(P_{ij} \ge \gamma)\Big)\\
\leq   &  (n - r + 1) \gamma \cdot \Pr(P_{(r - 1) j} < \gamma).
\end{align*}

\subsection{Proof of Theorem~\ref{thm:validity_two_step}}

First, for each $j = 1, 2, \dots, M$, we define
\[
	\gamma_j = \sup\Big\{\gamma \in 
    [0,\alpha]: 
    \gamma \cdot (1 + \sum_{s \neq j}^M 1_{F_s < \gamma}) \leq \alpha \Big\}.
\]
which is independent from $(F_j, S_j)$ under our independence assumption of the p-value matrix. It is obvious from the definition that we always have $\gamma_j \leq \gamma_0^{\text{Bon}}$. Specifically, if 
 $F_j < \gamma_0^{\text{Bon}}$, then $\gamma_j = \gamma_0^{\text{Bon}}$. 
%
%
Thus, as $F_j \leq S_j$ always holds, the PFER is
\begin{align*}
	\e(V) = \e\Big(\sum_{j = 1}^M 1_{S_j < \gamma_0^{\text{Bon}}} \cdot 1_{v_j = 0}\Big) 
       &= \e\Big(\sum_{j = 1}^M 1_{S_j < \gamma_0^{\text{Bon}}}1_{F_j < \gamma_0^{\text{Bon}}} 
  \cdot 1_{v_j = 0}\Big) \\
	& =  \sum_{j = 1}^M \e\Big(1_{S_j < \gamma_j}  \cdot 1_{F_j < \gamma_0^{\text{Bon}}}\cdot 1_{v_j = 0}\Big)\\
	& =  \sum_{j = 1}^M \e\Big(1_{S_j < \gamma_j}  \cdot 1_{F_j < \gamma_j}\Big) \cdot 1_{v_j = 0}
\end{align*}
The last equality holds as both $\gamma_j \leq \gamma_0^{\text{Bon}}$ and $F_j \leq S_j$ hold for any $j$.

Now using Lemma~\ref{lem:frac} and the fact that $\gamma_j \leq \gamma_0^{\text{Bon}}$, we have
\begin{align*}
	\e(V) 	
	& = \sum_{j = 1}^M \e\Big(1_{S_j < \gamma_j}  \cdot 1_{F_j < \gamma_j}\Big) \cdot 1_{v_j = 0} \\
	& =  \sum_{j = 1}^M \e
	\Bigg(\e\Big[1_{S_j < \gamma_j}  \mid \gamma_j, 1_{F_j < \gamma_j}\Big] 
	\cdot 1_{F_j < \gamma_j}\cdot 1_{v_j = 0}\Bigg)\\
	& \leq  \sum_{j = 1}^M\e\Big(\gamma_j 
	\cdot 1_{F_j < \gamma_j}\cdot 1_{v_j = 0}\Big) \\
	& \leq  \e\Big(\gamma_0^{\text{Bon}}  \cdot \sum_{j = 1}^M 
	1_{F_j < \gamma_0^{\text{Bon}}}\cdot 1_{v_j = 0}\Big)  \leq \alpha. 
\end{align*}
The last inequality holds as $\gamma_0^{\text{Bon}}$ itself satisfies $\gamma_0^{\text{Bon}}  \cdot \sum_{j = 1}^M 
	1_{F_j < \gamma_0^{\text{Bon}}} \leq \alpha$.

\subsection{Proof of  \Cref{thm:validity_FS_BH}}
For each $j = 1, 2, \dots, M$, define
\begin{align}\label{eq:defgammaj}
  \gamma_j = \sup\Big\{\gamma \in [0,\alpha]: 
    \gamma\cdot {(1 + \sum_{k \neq j^M} 1_{F_k < \gamma})}\leq \alpha \cdot 
  {(1 + \sum_{k \neq j}^M 1_{S_k < \gamma})}\Big\}.
\end{align}
which is independent from $(F_j, S_j)$. The relation between $\gamma_j$ and $\gamma_0^{\text{BH}}$ is complicated, and we discuss separately in different scenarios. First, if $S_j < \gamma_0^\text{BH}$, then as $F_j \leq S_j $, we have 
$\gamma_j \geq \gamma_0^{\text{BH}}$. On the other hand, since $S_j < \gamma_0^\text{BH} \leq \gamma_j$, we also have $\gamma_j \sum_k 1_{F_k < \gamma_j}\leq \alpha \sum_k 1_{S_k < \gamma_j}$ which indicates that $\gamma_j \leq \gamma_0^{\text{BH}}$. Thus, when $S_j < \gamma_0^\text{BH}$, we have $\gamma_j = \gamma_0^{\text{BH}}$. Second, if $F_j < \gamma_0^{\text{BH}} \le S_j$, then obviously $\gamma_j \geq \gamma_0^{\text{BH}}$. Finally, if $F_j \ge \gamma_0^{\text{BH}}$, since $\gamma_0^{\text{BH}} \leq \alpha$, we also have $\gamma_j \geq \gamma_0^{\text{BH}}$. 
In summary, $\gamma_j \geq \gamma_0^{\text{BH}}$ is always true, and when $S_j < \gamma_0^{\text{BH}}$, the equality holds.

Notice that for \Cref{def:two_step_MTP_BH}, the threshold $\gamma_0^{\text{BH}}$ itself satisfies 
$$\frac{\gamma_0^{\text{BH}} {\sum_{j = 1}^M 1_{F_j < \gamma_0^{\text{BH}}}}}{{\sum_{j = 1}^M 1_{S_j < \gamma_0^{\text{BH}}}} \vee 1}\leq \alpha $$
Thus for the FDR, we have
\begin{equation*}
  \begin{aligned}
    \e\left(\frac{V}{R \vee 1}\right)  = \e\left(\frac{\sum_{j = 1}^M 1_{S_j < \gamma_0^{\text{BH}}} 
\cdot 1_{v_j = 0}}{\sum_{j = 1}^M 1_{S_j < \gamma_0^{\text{BH}}}\vee 1}\right)
& \leq \alpha \cdot \e\left(\frac{\sum_{j = 1}^M 1_{S_j < \gamma_0^{\text{BH}}}\cdot 1_{v_j=0}
    }{[\gamma_0^{\text{BH}} \sum_{j = 1}^M 1_{F_j < \gamma_0^{\text{BH}}}] \vee \alpha }\right)\\
& = \alpha \cdot\sum_{j = 1}^M \e\left(\frac{1_{S_j < \gamma_0^{\text{BH}}}\cdot1_{v_j=0}}{[\gamma_0^{\text{BH}}\sum_{k = 1}^M 1_{F_k < \gamma_0^{\text{BH}}}] \vee \alpha}\right)\\
\end{aligned}
\end{equation*}
Making use of the relation between each $\gamma_j$ and $\gamma_0^{\text{BH}}$, we have for each $j$
\begin{equation*}
  \begin{aligned}
  \frac{1_{S_j < \gamma_0^{\text{BH}}}\cdot1_{v_j=0}}{[\gamma_0^{\text{BH}}\sum_{k = 1}^M 1_{F_k < \gamma_0^{\text{BH}}}] \vee \alpha}
  = \frac{1_{S_j < \gamma_0^{\text{BH}}} \cdot 1_{S_j < \gamma_j}\cdot1_{v_j=0}}{[\gamma_j(1 + \sum_{k \neq j}^M 1_{F_k < \gamma_j})] \vee \alpha}
\leq \frac{1_{S_j < \gamma_j}\cdot 1_{v_j=0}}{\big[\gamma_j (1 + \sum_{k \neq j}^M 1_{F_k < \gamma_j})\big]\vee \alpha}.
\end{aligned}
\end{equation*}

Now let $P_{\cdot (-j)}$ contain all $P_{\cdot k}$ for $k\ne j$. This $P_{\cdot(-j)}$   
 determines $\gamma_j$ and all $F_k$ for $k\ne j$. 
Combing the last two steps, as $\gamma_j$ is independent from $(F_j, S_j)$, using Lemma~\ref{lem:frac}, we have
\begin{align*}
\e\left(\frac{V}{R \vee 1}\right) 
&\le \alpha \cdot\sum_{j = 1}^M \e\left(\e\left[\frac{1_{S_j < \gamma_j}\cdot1_{v_j=0}}{\big[\gamma_j(1 + \sum_{k \neq j}^M 1_{F_k < \gamma_j})\big] \vee \alpha}\Bigm|P_{\cdot(-j)}\right]\right)\\
&\le \alpha \cdot\sum_{j = 1}^M \e\left(\frac{\gamma_j\e[1_{F_j < \gamma_j}\mid P_{\cdot(-j)}]}{ \big[\gamma_j (1 + \sum_{k \neq j}^M 1_{F_k < \gamma_j})\big] \vee \alpha}\right)\\
&= \alpha \cdot\sum_{j = 1}^M \e\left(\frac{\gamma_j1_{F_j < \gamma_j}}{\big[\gamma_j(1 + \sum_{k \neq j}^M 1_{F_k < \gamma_j})\big] \vee \alpha}\right).
\end{align*}
Next, because $1_{F_j<\gamma_j}\le1$ and $\gamma_j\le\alpha$,
\begin{align*}
\e\left(\frac{V}{R \vee 1}\right) 
&\le \alpha \cdot\sum_{j = 1}^M \e\left(\frac{\gamma_j1_{F_j < \gamma_j}}{\big[\gamma_j \sum_{k=1}^M 1_{F_k < \gamma_j}\big] \vee \alpha}\right)
\le\alpha\sum_{j=1}^M\e\left(
\frac{1_{F_j<\gamma_j}}{\sum_{k=1}^M1_{F_k<\gamma_j} \vee 1}
\right).
\end{align*}
WLOG, we can assume $F_1\le F_2\le\cdots\le F_M$. To complete the proof, note that no matter $F_j<\gamma_j$ or not, we always have
$$
\frac{1_{F_j<\gamma_j}}{\sum_{k=1}^M1_{F_k<\gamma_j} \vee 1}
\le\frac{1_{F_j<\gamma_j}}{\sum_{k=1}^M1_{F_k\le F_j}}
\le\frac{1}{\sum_{k=1}^M1_{F_k\le  F_j}}
\le\frac1j.
$$
Thus,
$$\e\left(\frac{V}{R \vee 1}\right)  \leq \alpha \sum_j \frac{1}{j}.$$

\subsection{Proof of  \Cref{thm:BH_asymptotic_dependence}}
We separate the proof into three parts. The first part shows convergence of some empirical cumulative distribution functions (ECDFs).  Then the next two parts establish the two claims in the theorem.


The first part of that proof requires weak dependence of the filtered $p$-values $F_j$.  Our next lemma extends weak dependence from the base $p$-values $P_{ij}$ to the $F_j$.
\begin{lemma}\label{lem:tech}
\Cref{assp:weak} guarantees that for any fixed $\gamma$,
$$\frac{1}{M^2}\sum_{j\neq j'}\big|\Pr(F_{j}<\gamma,F_{j'}<\gamma) - \Pr(F_{j}<\gamma)\Pr(F_{j'}<\gamma)\big|\overset{M \to \infty}{\longrightarrow}0$$
\end{lemma}


\begin{proof}
Because $F_j  = (n - r + 1)P_{(r-1)j}$, we only need to show that
$$A = \frac{1}{M^2} \sum_{j\neq j'}\left[\Pr(P_{(r-1)j}<\gamma,\ P_{(r-1)j'}<\gamma)- \Pr(P_{(r-1)j}<\gamma)\Pr(P_{(r-1)j'}<\gamma)\right] \to 0.$$
With the following decomposition:
\begin{align*}
&\Pr(P_{(r-1)j}<\gamma,\ P_{(r-1)j'}<\gamma)\\
&= \sum_{k=r-1}^n\sum_{\tilde{k}=r-1}^n\sum_{\mbox{\tiny{$\begin{array}{c}
     u\subset1{:}n\\
     |u|=k 
\end{array}$}}}
\sum_{\mbox{\tiny{$\begin{array}{c}
     \tilde{u} \subset 1{:}n \\
     |\tilde{u}|=\tilde{k}
\end{array}$}}} \Pr\left(\bm P_{uj}\prec \gamma, \bm P_{u^cj} \succcurlyeq \gamma,
\bm P_{\tilde{u}j'}\prec \gamma, \bm P_{\tilde{u}^cj'} \succcurlyeq \gamma\right)
\end{align*}
and
\begin{align*}
&\Pr(P_{(r-1)j}<\gamma)\Pr(\ P_{(r-1)j'}<\gamma)\\
&= \sum_{k=r-1}^n\sum_{\tilde{k}=r-1}^n\sum_{\mbox{\tiny{$\begin{array}{c}
     u\subset1{:}n\\
     |u|=k 
\end{array}$}}}
\sum_{\mbox{\tiny{$\begin{array}{c}
     \tilde{u} \subset 1{:}n \\
     |\tilde{u}|=\tilde{k}
\end{array}$}}} \Pr\left(\bm P_{uj}\prec \gamma, \bm P_{u^cj} \succcurlyeq \gamma)\Pr(
\bm P_{\tilde{u}j'}\prec \gamma, \bm P_{\tilde{u}^cj'} \succcurlyeq \gamma\right),
\end{align*}
we further only need to show that for any sets $u, \tilde{u} \subset 1{:}n$ and any $j\neq j' \in 1{:}M$, 
\begin{align*}
\Delta_{u,\tilde u, jj'} = & |\Pr\left(\bm P_{uj}\prec \gamma, \bm P_{u^cj} \succcurlyeq \gamma,
\bm P_{\tilde{u}j'}\prec \gamma, \bm P_{\tilde{u}^cj'} \succcurlyeq \gamma\right) \\
& -\Pr\left(\bm P_{uj}\prec \gamma, \bm P_{u^cj} \succcurlyeq \gamma)\Pr(
\bm P_{\tilde{u}j'}\prec \gamma, \bm P_{\tilde{u}^cj'} \succcurlyeq \gamma\right)| 
\end{align*}
converges to $0$ when $M \to \infty$ and $n$ is fixed. Since we assume that base p-values across studies are 
independent, 
we have
\begin{align*}
    &\Pr\left(\bm P_{uj}\prec \gamma, \bm P_{u^cj}\succcurlyeq \gamma,
\bm P_{\tilde{u}j'}\prec \gamma, \bm P_{\tilde{u}^cj'}\succcurlyeq \gamma\right)\\
=& \prod_{t\in u-\tilde{u}}\Pr(P_{tj}<\gamma) 
\prod_{\tilde{t}\in\tilde{u}-u} \Pr(P_{\tilde{t}j'}<\gamma)
\prod_{o\in u^c-\tilde{u}^c} \Pr(P_{oj}\ge\gamma)\prod_{\tilde{o}\in\tilde{u}^c-u^c} 
\Pr(P_{\tilde{o}j'}\ge\gamma) \\
& \times \prod_{t'\in u\cap\tilde{u}} \Pr(P_{t'j}<\gamma,\ P_{t'j'}<\gamma)
\prod_{o'\in u^c\cap\tilde{u}^c} \Pr(P_{o'j}\ge\gamma,\ P_{o'j'}\ge\gamma)
\end{align*}
and for $s = j$ or $j'$ and any $u$,
\begin{align*}
  \Pr\left(\bm P_{us}\prec \gamma, \bm P_{u^cs} \succcurlyeq \gamma\right) = 
  \prod_{l\in u}\Pr(P_{ls}<\gamma)\prod_{h\in u^c}\Pr(P_{hs}\ge\gamma).
\end{align*}
Thus, after merging the shared probability terms  and bounding them by $1$, we have
\begin{align*}
\Delta_{u,\tilde u, jj'}
\leq & \left|\prod_{t'\in u\cap\tilde{u}} \Pr(P_{t'j}<\gamma,\ P_{t'j'}<\gamma)
\prod_{o'\in u^c\cap\tilde{u}^c} \Pr(P_{o'j}\ge\gamma,\ P_{o'j'}\ge\gamma) \right.\\
& \left.-\prod_{t'\in u\cap\tilde{u}} \Pr(P_{t'j}<\gamma)\Pr(P_{t'j'}<\gamma)
\prod_{o'\in u^c\cap\tilde{u}^c} \Pr(P_{o'j}\ge\gamma)\Pr(P_{o'j'}\ge\gamma)\right|.
\end{align*}
Next for any $a_k,b_k\in[0,1]$ we have $|\prod_{k=1}^na_k-\prod_{k=1}^nb_k|\le \sum_{k=1}^n|a_k-b_k|$.
From this inequality and \Cref{assp:weak},
\begin{align*}
\Delta_{u,\tilde u, jj'}
\leq & \sum_{t'\in u\cap\tilde{u}} \Big|\Pr(P_{t'j}<\gamma,\ P_{t'j'}<\gamma)
-\Pr(P_{t'j}<\gamma)\Pr(P_{t'j'}<\gamma)\Big|\\
&+
\sum_{o'\in u^c\cap\tilde{u}^c} \Big|\Pr(P_{o'j}\ge\gamma,\ P_{o'j'}\ge\gamma)- \Pr(P_{o'j}\ge\gamma)\Pr(P_{o'j'}\ge\gamma)\Big|\\
\leq& \sum_{l=1}^n \big|\Pr(P_{lj}<\gamma,\ P_{lj'}<\gamma)-\Pr(P_{lj}<\gamma)\Pr(P_{lj'}<\gamma)\big| \to 0
\end{align*}
when $M \to 0$. Thus, $A\to \infty$ and the Lemma is proved.
\end{proof}
Now we are ready to prove the three parts.

\vspace{2mm}\noindent
\noindent\textbf{PART I: ECDF convergence.}

\vspace{2mm}\noindent Define four ECDFs
\begin{align*}
{F_{0,M}(\gamma)} &:=
 \frac{1}{M_0}\sum_{j\in\mathcal{H}_0^{r/n}}1_{F_j<\gamma},\ \ \ 
&{F_{1,M}(\gamma)} &:= \frac{1}{M_1}\sum_{j\in\mathcal{H}_1^{r/n}}1_{F_j<\gamma}\\
{S_{0,M}(\gamma)} &:=
 \frac{1}{M_0}\sum_{j\in\mathcal{H}_0^{r/n}}1_{S_j<\gamma},\quad \text{and}
&{S_{1,M}(\gamma)} &:= \frac{1}{M_1}\sum_{j\in\mathcal{H}_1^{r/n}}1_{S_j<\gamma}.
\end{align*}
We will show that $F_{0,M}(\gamma)\overset{p}\to \tilde F_0(\gamma)$ uniformly in $0\le\gamma\le1$ under Assumptions \ref{assp:weak} and \ref{assp:asymptotic}.  The same argument  establishes uniform convergence of $F_{1,M}\overset{p}\to\tilde F_{1}$, $S_{0,M}\overset{p}\to\tilde S_{0}$ and $S_{1,M}\overset{p}\to\tilde S_{1}$.

First we write
\begin{align}\label{eq:ecdfdecomp}
|F_{0,M}(\gamma)-\tilde F_{0}(\gamma)|
\le|F_{0,M}(\gamma)-\e(F_{0,M}(\gamma))|+|\e(F_{0,M}(\gamma))-\tilde F_{0}(\gamma)|.
\end{align}
The second term in~\eqref{eq:ecdfdecomp} vanishes pointwise in $\gamma$ by \Cref{assp:asymptotic}. 
Next by \Cref{lem:tech},
$$
\mathrm{var}(F_{0,M}(\gamma)) = \frac{\sum_{j\in\mathcal{H}_0^{r/n}}\Pr(F_j<\gamma)\Pr(F_j\ge\gamma)}{M_0^2} + o(1)\to0
$$
and so by Chebychev's inequality, the first term in~\eqref{eq:ecdfdecomp} also vanishes pointwise in $\gamma$.  This proves pointwise convergence of $F_{0,M}$ to $\tilde F_0$.  Then uniform convergence follows the same way it does in the Glivenko-Cantelli theorem.

\vspace{6mm}\noindent \textbf{PART II: Proof of $\gamma_0^\text{BH}\to\gamma_0^\infty$.}

\vspace{2mm}\noindent
Define
$$ {F_{M}(\gamma)}:=\frac{1}{M}\sum_{i=1}^M 1_{(F_i<\gamma)}, \ \
{S_{M}(\gamma)}:=\frac{1}{M}\sum_{i=1}^M 1_{(S_i<\gamma)}
\quad \text{and}\quad 
{f_M(\gamma)} := \frac{\gamma F_M(\gamma)}{S_M(\gamma)\vee\frac{1}{M}}$$ 
A direct conclusion from Part I is that 
\begin{align*}
F_{M}(\gamma)\overset{p}{\to}\tilde{F}(\gamma)&,\quad 
S_{M}(\gamma)\overset{p}{\to}\tilde{S}(\gamma)\quad\text{and}\quad
f_M(\gamma)\overset{p}{\to}f^\infty(\gamma)
\end{align*}
all hold uniformly in $\gamma\in[0,\alpha]$. 

As a consequence, for $\forall x \in [0, \alpha]$,
\begin{align*}
\inf_{\gamma\in[x,\alpha] }f_M(\gamma)\ge&\inf_{\gamma\in[x,\alpha]}\left[f_M(\gamma)-f^\infty(\gamma)\right]+\inf_{\gamma\in[x,\alpha]}f^\infty(\gamma)\\
\ge& -\sup_{\gamma\in[x,\alpha]}\left|f_M(\gamma)-f^\infty(\gamma)\right|+\inf_{\gamma\in[x,\alpha]}f^\infty(\gamma)\\
\overset{p}{\to}& \inf_{\gamma\in[x,\alpha]}f^\infty(\gamma).
\end{align*}
In addition, for any $\epsilon > 0$ we have
\begin{align*}
    & \limsup_{M \to \infty} \Pr(\inf_{\gamma\in[x,\alpha] }f_M(\gamma)\leq \alpha)  \\
    \leq &
\Pr\left(\liminf_{M \to \infty}(\inf_{\gamma\in[x,\alpha] }f_M(\gamma)) - \inf_{\gamma\in[x,\alpha] }f^\infty(\gamma) < -\epsilon\right) + 
1_{\inf_{\gamma\in[x,\alpha] }f^\infty(\gamma)\leq \alpha + \epsilon}
\end{align*}
Let $\epsilon \to 0$, we have
$$\limsup_{M \to \infty} \Pr(\inf_{\gamma\in[x,\alpha] }f_M(\gamma)\leq \alpha) \leq 1_{\inf_{\gamma\in[x,\alpha] }f^\infty(\gamma)\leq \alpha}$$
Similarly,
\begin{align*}
\inf_{\gamma\in[x,\alpha]}f_M(\gamma)\le & \sup_{\gamma\in[x,\alpha]}\left|f_M(\gamma)-f^\infty(\gamma)\right|+\inf_{\gamma\in[x,\alpha]}f^\infty(\gamma)\\
\overset{p}{\to}& \inf_{\gamma\in[x,\alpha]}f^\infty(\gamma).
\end{align*}
and we have
$$\liminf_{M \to \infty} \Pr(\inf_{\gamma\in[x,\alpha] }f_M(\gamma)\leq \alpha) \geq \lim_{\epsilon \to 0}1_{\inf_{\gamma\in[x,\alpha] }f^\infty(\gamma)\leq \alpha - \epsilon} = 1_{\inf_{\gamma\in[x,\alpha] }f^\infty(\gamma) < \alpha}$$

Notice that
\begin{align*}
E[{(\gamma_0^\text{BH})}^k]=&\int_0^\alpha kx^{k-1}P(\gamma_0^{\text{BH}}\ge x)dx\\
=&\int_0^\alpha kx^{k-1}P(\inf_{\gamma\in[x,\alpha] }f_M(\gamma)\le\alpha)dx,
\end{align*}
Thus, by Fatou's lemma
\begin{align*}
\int_0^\alpha kx^{k-1}1_{\inf_{\gamma\in[x,\alpha] }f^\infty(\gamma) < \alpha}dx
\le&\liminf_{M\to\infty}E[{(\gamma_0^\text{BH})}^k]\\
\le&\limsup_{M\to\infty}E[{(\gamma_0^\text{BH})}^k]\\
\le&\int_0^\alpha kx^{k-1}1_{\inf_{\gamma\in[x,\alpha] }f^\infty(\gamma) \leq \alpha}dx.
\end{align*}
In addition, we have $\tilde F(\gamma) \geq \tilde S(\gamma)$ as $F_j \leq S_j$ for any hypothesis $j$, thus
$\gamma_0^\infty \leq \alpha$
and then \Cref{assp:technical}(a) and $f^\infty$'s left continuity also guarantee that 
$$\{x:\ \gamma_0^\infty \geq x \text{ and } x\leq \alpha\} = \left\{x:\ \inf_{\gamma\in[x,\alpha]}f^\infty(\gamma)\le\alpha\right\}=
\left\{x:\ \inf_{\gamma\in[x,\alpha]}f^\infty(\gamma)<\alpha\right\} \cup \{\gamma_0^\infty\}.$$
Hence,
\begin{align*}
\int_0^\alpha kx^{k-1}1_{\inf_{\gamma\in[x,\alpha] }f^\infty(\gamma) < \alpha}dx=&
\int_0^\alpha kx^{k-1}1_{\inf_{\gamma\in[x,\alpha] }f^\infty(\gamma) \leq \alpha}dx\\
=&\int_0^{\gamma_0^\infty}kx^{k-1}dx\\
=&{(\gamma_0^\infty)}^k.
\end{align*}
Then
\begin{align*}
& E|\gamma_0^\text{BH}-\gamma_0^\infty|^2=E[{(\gamma_0^\text{BH})}^2]-2\gamma_0^\infty\cdot E[\gamma_0^\text{BH}]+{(\gamma_0^\infty)}^2\rightarrow0\\
\Rightarrow& P(|\gamma_0^\text{BH}-\gamma_0^\infty|\ge\epsilon)\le\frac{E|\gamma_0^\text{BH}-\gamma_0^\infty|^2}{\epsilon^2}\rightarrow0\\
\Rightarrow&\gamma_0^\text{BH}\overset{p}{\to}\gamma_0^\infty
\end{align*}
where $\gamma_0^\infty$ is a constant.

\vspace{6mm}\noindent \textbf{PART III: Convergence of FDP.}

\vspace{4mm}\noindent Finally, we prove that if $\tilde{S}(\gamma_0^\infty)>0$, then
$$\text{FDP}\overset{p}{\to}\frac{\pi_0 \tilde S_0(\gamma_0^\infty)}{\tilde{S}(\gamma_0^\infty)}\ \le\alpha$$
where $\text{FDP}$ is the false discovery proportion of the AdaFilter BH procedure.

Since we have already shown in Part II that 
$$\gamma_0^\text{BH}\overset{p}{\to}\gamma_0^\infty,$$
then $\forall\epsilon\in(0,\gamma_0^\infty)$ and  $\forall\eta\in(0,1)$,
$$\frac{M_0S_{0,M}(\gamma_0^\infty-\epsilon)/M}{S_M(\gamma_0^\infty+\epsilon)\vee\frac{1}{M}}\le \frac{M_0S_{0,M}(\gamma_0^{\text{BH}})}{MS_M(\gamma_0^\text{BH})\vee1}\le\frac{M_0S_{0,M}(\gamma_0^\infty+\epsilon)/M}{S_M(\gamma_0^\infty-\epsilon)\vee \frac{1}{M}}$$
hold with probability at least $1-\eta$ when $M$ is sufficiently large.


Then as $M_0/M\to\pi_0$, $S_{0,M}(\gamma)\overset{p}{\to}\tilde S_0(\gamma)$, $S_M(\gamma)\overset{p}{\to}\tilde{S}(\gamma)$ uniformly in $\gamma\in[0,1]$, and $\tilde S_0$, $\tilde S_1$ are continuous at $\gamma_0^\infty$ by \Cref{assp:technical}(b), so letting $\epsilon\to0$, we can easily get,
$$\text{FDP}\overset{p}{\to}\frac{\pi_0 \tilde S_0(\gamma_0^\infty)}{\tilde{S}(\gamma_0^\infty)}\le \frac{\pi_0 \gamma_0^\infty \tilde F_0(\gamma_0^\infty)}{\tilde{S}(\gamma_0^\infty)}
\le\frac{\gamma_0^\infty\tilde{F}(\gamma_0^\infty)}{\tilde{S}(\gamma_0^\infty)}\le\alpha.$$
The first inequality is due to \Cref{lem:frac}, the second inequality is because of the continuity assumptions in \Cref{assp:technical}(a), and the last inequality is by the definition of $\gamma_0^\infty$ and the left-continuity of $f^{\infty}$. 






\subsection{Proof of Corollary~\ref{cor:mono}}

For some $j$, let $\tilde p_{\cdot j} = (\tilde p_{1j}, \cdots, \tilde p_{n j})$ 
satisfy $\tilde p_{ij} \leq p_{ij}$ for $i = 1, 2, \cdots, n$.  Now construct
a new $N\times n$ $P$-value matrix $\tilde P$ with the given row $\tilde P_{\cdot j}$
and all other rows $\tilde P_{\cdot k} = P_{\cdot k}$ for $k\ne j$.
Define $(\tilde F_{1}, \cdots, \tilde F_{M})$ as the 
corresponding filtering statistics \eqref{eq:filterp} and $(\tilde S_{1}, \cdots, \tilde S_{M})$ as the 
corresponding selection statistics \eqref{eq:selectp} with $\tilde P$ replacing $P$. 
Then $\tilde F_k = F_k$ and $\tilde S_k = S_k$ for $k \neq j$ and $\tilde F_j \leq F_j$ with $\tilde S_j \leq S_j$.

For the AdaFilter Bonferroni procedure, let $\tilde \gamma_j^{\text{Bon}}$ 
be the new $\gamma_0^{\text{Bon}}$ using the new base P-values. 
For the AdaFilter BH procedure, let $\tilde \gamma_j^{\text{BH}}$ 
be the new $\gamma_0^{\text{BH}}$ using the new base p-values. 
Then to show that the procedures satisfy partial monotonicity, we only need to show that if 
$S_j < \gamma_0$, then $\tilde S_j < \tilde\gamma_j$ for both the Bonferroni correction and BH.

For the AdaFilter Bonferroni procedure, if $S_j < \gamma_0$, then $\tilde S_j < \gamma_0$, thus 
$$\gamma_0^{\text{Bon}}\cdot \sum_{k = 1}^M 1_{\tilde F_k<  \gamma_0^{\text{Bon}}} \leq \alpha
$$
which means that $\tilde \gamma_j^{\text{Bon}} \geq \gamma_0^{\text{Bon}}$. 
Similarly, for the AdaFilter BH procedure using the same argument, we have 
$\tilde \gamma_j^{\text{BH}} \geq \gamma_0^{\text{BH}}$ when $S_j < \gamma_0$.
As a consequence, for both AdaFilter procedures, we have 
$\tilde S_j \leq S_j < \gamma_0 \leq \tilde \gamma_j$.

\begin{table}[ht]
\centering
\scalebox{0.95} {
\begin{tabular}{@{}lrr@{}}
  \hline
 & $P_{4/4}$ & GO Biological Process \\ 
  \hline
MYH3 & 3.31e-07 & actin filament-based movement, muscle organ development, \\
&& striated muscle contraction \\
  S100A4 & 1.11e-06 & epithelial to mesenchymal transition \\ 
  S100A10 & 1.50e-04 & signal transduction, regulation of cell growth, regulation of cell differentiation \\ 
  S100A13 & 2.87e-04 & cell differentiation \\ 
  TMSB10 & 7.45e-04 & sequestering of actin monomers, actin cytoskeleton organization \\ 
  PLAU & 1.15e-03 & angiogenesis, fibrinolysis, signal transduction, regulation of cell proliferation, 
  \\ && blood coagulation, smooth muscle cell migration, embryo implantation, \\ && skeletal muscle regeneration, chemotaxis $<$more data available...$>$ \\ 
  CLIC1 & 1.33e-03 & signal transduction, transport, ion transport, chloride transport, \\ && response to unfolded protein, response to nutrient, defense response \\ 
  PLA2G2A & 1.64e-03 & negative regulation of cell proliferation, somatic stem cell maintenance, negative \\ && regulation of epithelial cell proliferation, positive regulation of foam cell \\ &&
   differentiation,  positive regulation of inflammatory response \\ && $<$more data available...$>$ \\ 
  MYL5 & 1.87e-03 & regulation of muscle contraction \\ 
  CHRNA1 & 2.11e-03 & signal transduction, muscle maintenance, neuron maintenance, regulation of \\ && membrane potential, regulation of action potential in neuron, neuromuscular \\&&  synaptic transmission,  neuromuscular 
 junction development, skeletal muscle tissue  \\ && growth, transport, ion transport $<$more data available...$>$ \\ 
  TYROBP & 2.31e-03 & intracellular signaling cascade, cellular defense response \\ 
  ART3 & 2.46e-03 & protein amino acid ADP-ribosylation \\ 
  MYH8 & 2.58e-03 & biological\_process, striated muscle contraction \\ 
  DAB2 & 3.37e-03 & negative regulation of cell growth, cell proliferation, cell morphogenesis involved in\\ &&differentiation, pinocytosis, receptor-mediated endocytosis, in utero embryonic \\&&  development, excretion \\ 
  S100A11 & 3.71e-03 & negative regulation of cell proliferation, negative regulation of DNA replication, \\ && signal transduction \\ 
  LAPTM5 & 4.48e-03 & transport \\ 
  EEF1A1 & 4.59e-03 & translational elongation \\ 
  IGFBP4 & 4.82e-03 & signal transduction, regulation of cell growth, intracellular signaling cascade, 
  \\&& cell proliferation, DNA metabolic process, skeletal system development, \\&& inflammatory response \\ 
  TUBA1A & 7.97e-03 & protein polymerization, microtubule-based process, microtubule-based movement \\ 
  F13A1 & 8.43e-03 & blood coagulation, peptide cross-linking, wound healing \\ 
  RPL3 & 9.97e-03 & biological\_process, translation, translational elongation \\ 
  ANXA2 & 1.10e-02 & angiogenesis, fibrinolysis, collagen fibril organization \\ 
  PPP1R1A & 1.20e-02 & signal transduction, glycogen metabolic process, carbohydrate metabolic process \\ 
  MYL4 & 1.21e-02 & regulation of the force of heart contraction, positive regulation of ATPase activity, \\ && muscle organ development, cardiac muscle contraction \\ 
  SRPX & 1.28e-02 & biological\_process, cell adhesion \\ 
  HLA-DPB1 & 1.44e-02 & immune response, antigen processing and presentation of peptide or\\ &&  polysaccharide antigen via MHC class II \\ 
   \hline
\end{tabular}
}
\caption{$27$ or the $32$ AdaFilter selected genes for $r = 4$ with FDR controlled at $\alpha = 0.05$ where functional annotations are available in \cite{kotelnikova2012}(Table S2). The AdaFilter selection p-values for theses genes are also reported.}
\label{tab:gwas_result}
\end{table}

\begin{table}[ht]
\centering
\scalebox{1}{
\begin{tabular}{l|cccc}
\hline
\multicolumn{1}{c|}{\textbf{Gene}} & \bm{$F_j$} & \bm{$S_j$} & \textbf{$m_j^{\text{AF}}$}  & \bm{$p_j^{\text{BH}}$}  \\ \hline
\textbf{TNFRSF18}                  & 1.50e-04          & 1.97e-04          & 1                     & 1.97e-04              \\
\textbf{GAPDH}                     & 1.90e-03          & 2.63e-03          & 9                     & 6.52e-03              \\
\textbf{SRGN}                      & 8.63e-04          & 2.74e-03          & 9                     & 6.52e-03              \\
\textbf{CD7}                       & 2.10e-03          & 2.90e-03          & 9                     & 6.52e-03              \\
\textbf{BATF}                      & 3.48e-03          & 4.10e-03          & 11                    & 9.03e-03              \\
\textbf{DUSP4}                     & 1.68e-03          & 4.90e-03          & 14                    & 1.14e-02              \\
\textbf{CCR8}                      & 3.03e-03          & 5.34e-03          & 15                    & 1.14e-02              \\
\textbf{LAYN}                      & 4.69e-03          & 6.93e-03          & 18                    & 1.43e-02              \\
\textbf{TNFRSF9}                   & 4.61e-03          & 7.15e-03          & 18                    & 1.43e-02              \\
\textbf{SDC4}                      & 5.88e-03          & 9.99e-03          & 19                    & 1.90e-02              \\
\textbf{TNFRSF4}                   & 1.06e-02          & 1.19e-02          & 20                    & 2.16e-02              \\
\textbf{PHTF2}                     & 5.21e-03          & 1.31e-02          & 21                    & 2.30e-02              \\
\textbf{CXCR6}                     & 1.81e-03          & 1.56e-02          & 24                    & 2.87e-02              \\
\textbf{CREM}                      & 5.43e-03          & 1.57e-02          & 26                    & 2.92e-02              \\
\textbf{TNFAIP3}                   & 1.41e-03          & 2.08e-02          & 28                    & 3.89e-02              \\
\textbf{LYST}                      & 1.26e-02          & 2.35e-02          & 30                    & 4.34e-02              \\
\textbf{PHLDA1}                    & 2.20e-03          & 2.41e-02          & 31                    & 4.34e-02              \\
\textbf{CTLA4}                     & 1.51e-02          & 2.54e-02          & 31                    & 4.34e-02              \\
\textbf{FOXP3}                     & 4.57e-03          & 2.66e-02          & 31                    & 4.34e-02              \\
\textbf{ID2}                       & 2.15e-02          & 2.76e-02          & 32                    & 4.41e-02                 \\
\hline
\end{tabular}
}
\caption{scRNA-seq data analysis: marker genes at $r = 10$. Filtering, selection and Bonferroni BH adjusted p-values for the rejected marker genes at $r =10$ are shown, as long as the AdaFilter adjustment number. FDR controlled at $\alpha = 0.05$.}
\label{tab:sc_marker_genes}
\end{table}

\begin{table}[ht]
\centering
\scalebox{0.85}{
\begin{tabular}{@{}llrrrrrrrr@{}}
  \hline
 SNP & Gene & Amino acid & Carbohydrate & Cofactors  & Energy & Lipid & Nucleotide & Peptide & Xenobiotics \\ &&&&and vitamins&&&&& \\
  \hline
rs780093 & GCKR & \textbf{ 1.1e-25} & \textbf{7.4e-121} &  \textbf{ 2.8e-07} &  2.0e-02 & \textbf{ 5.8e-20} &  2.1e-02 &  5.8e-03 &  9.9e-01 \\ 
  rs6430538 & CCNT2-AS1 & \textbf{1.3e-04} & \textbf{9.5e-06} & 2.3e-02 & 5.4e-01 & 4.5e-01 & \textbf{6.8e-05} & 1.8e-01 & 4.4e-01 \\ 
  rs715 & CPS1 & \textbf{1.4e-228} &  5.5e-02 &  3.6e-02 &  \textbf{1.5e-03} & \textbf{ 3.0e-08} &  2.0e-01 & \textbf{ 3.5e-04} &  5.1e-01 \\ 
  rs6449202 & SLC2A9 & \textbf{1.3e-05} & 1.3e-01 & \textbf{1.5e-04} & 5.3e-01 & 2.9e-01 & \textbf{3.7e-26} & 3.4e-01 & 3.9e-01 \\ 
  rs11950562 & SLC22A4 & \textbf{2.6e-34} & 7.3e-01 & 7.5e-02 & 2.0e-01 & \textbf{6.9e-16} & 3.2e-01 & 8.1e-01 & \textbf{6.2e-05} \\ 
  rs4074995 & RGS14 & \textbf{4.7e-04} & 6.3e-03 & \textbf{5.2e-04} & 4.6e-01 & 9.2e-02 & 1.0e-02 & \textbf{4.4e-05} & 1.6e-01 \\ 
  rs1179087 & SLC17A3 & \textbf{1.0e-03} & 6.1e-01 & \textbf{3.2e-04} & 5.7e-01 & \textbf{2.6e-04} & \textbf{4.6e-04} & 1.7e-01 & 2.7e-01 \\ 
  rs657152 & ABO & \textbf{1.2e-04} & 5.7e-03 & 8.7e-02 & \textbf{1.0e-04} & 1.4e-01 & 2.0e-03 & \textbf{1.3e-26} & 1.1e-01 \\ 
  rs964184 & ZPR1 & 4.6e-02 & \textbf{8.2e-05} & \textbf{6.9e-04} & 6.4e-03 & \textbf{9.1e-20} & 5.0e-02 & 1.1e-01 & 6.5e-01 \\ 
  rs11045819 & SLCO1B1 & \textbf{8.5e-04} & 8.0e-01 & \textbf{3.6e-04} & 1.5e-01 & \textbf{2.0e-38} & 6.9e-02 & 9.1e-01 & 1.7e-02 \\ 
  rs4149056 & SLCO1B1 & \textbf{ 3.1e-04} &  5.6e-01 & \textbf{ 4.8e-04} &  8.3e-01 & \textbf{3.3e-142} &  2.4e-01 &  6.5e-01 &  8.4e-01 \\ 
  rs2062541 & ABCC1 & \textbf{7.1e-05} & 2.1e-01 & 8.8e-01 & \textbf{1.7e-14} & \textbf{2.9e-08} & 4.4e-01 & \textbf{1.1e-06} & 9.0e-01 \\ 
  rs310331 & ZNF19 & \textbf{5.9e-04} & 9.7e-01 & \textbf{6.4e-04} & 5.9e-01 & 3.9e-01 & 1.0e-01 & \textbf{1.3e-05} & 1.4e-01 \\ 
  rs7225637 & CCDC57 & \textbf{3.1e-10} & 8.7e-02 & 6.7e-02 & \textbf{4.2e-06} & \textbf{1.3e-05} & 8.3e-01 & 9.4e-01 & 4.4e-01 \\ 
   \hline
\end{tabular}
}
\caption{Metabalics GWAS data analysis: significant SNPs at $r = 3$ after clumping ($r^2$ set to 0.1 in PLINK).  Base  p-value for each of the 8 super-pathways are shown. The significant ones for each marker are in bold. FDR controlled at $\alpha = 0.05$.}
\label{tab:meta_result}
\end{table}

\newpage

\begin{figure}
  \centering
\includegraphics[width = 0.85\textwidth]{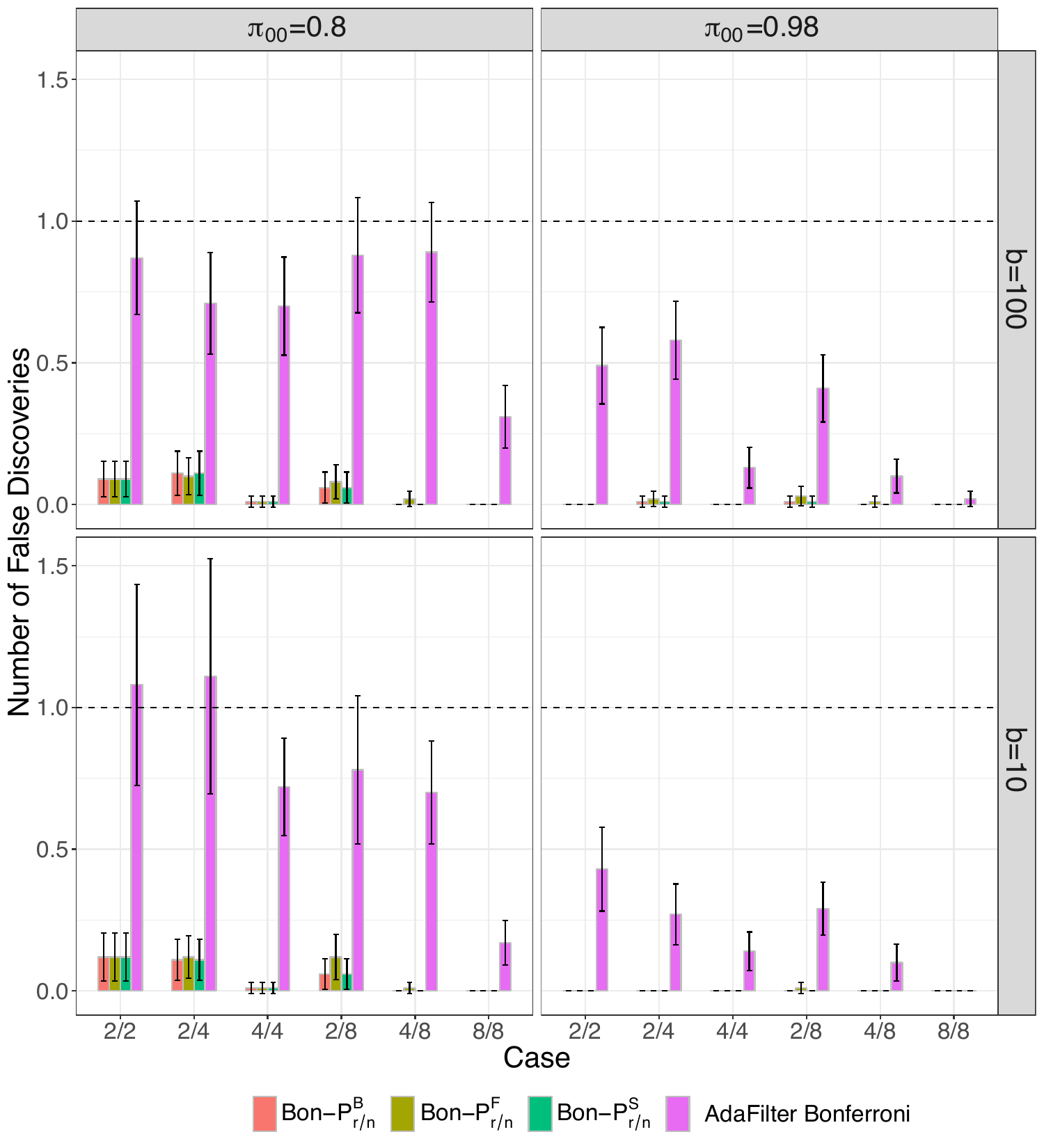}
\caption{Comparison of expected number of false discoveries $E(V)$ (PFER). The 
dotted line indicates the nominal level $\alpha = 1$. The error bars are the $95\%$ 
CI of the PFER from $B = 100$ experiments. The number of blocks $b = 100$ is the weak dependence scenario and $b = 10$ is the strong dependence scenario.}
\label{fig:PFER_two}
\end{figure}

\begin{figure}
  \centering
\includegraphics[width = 0.85\textwidth]{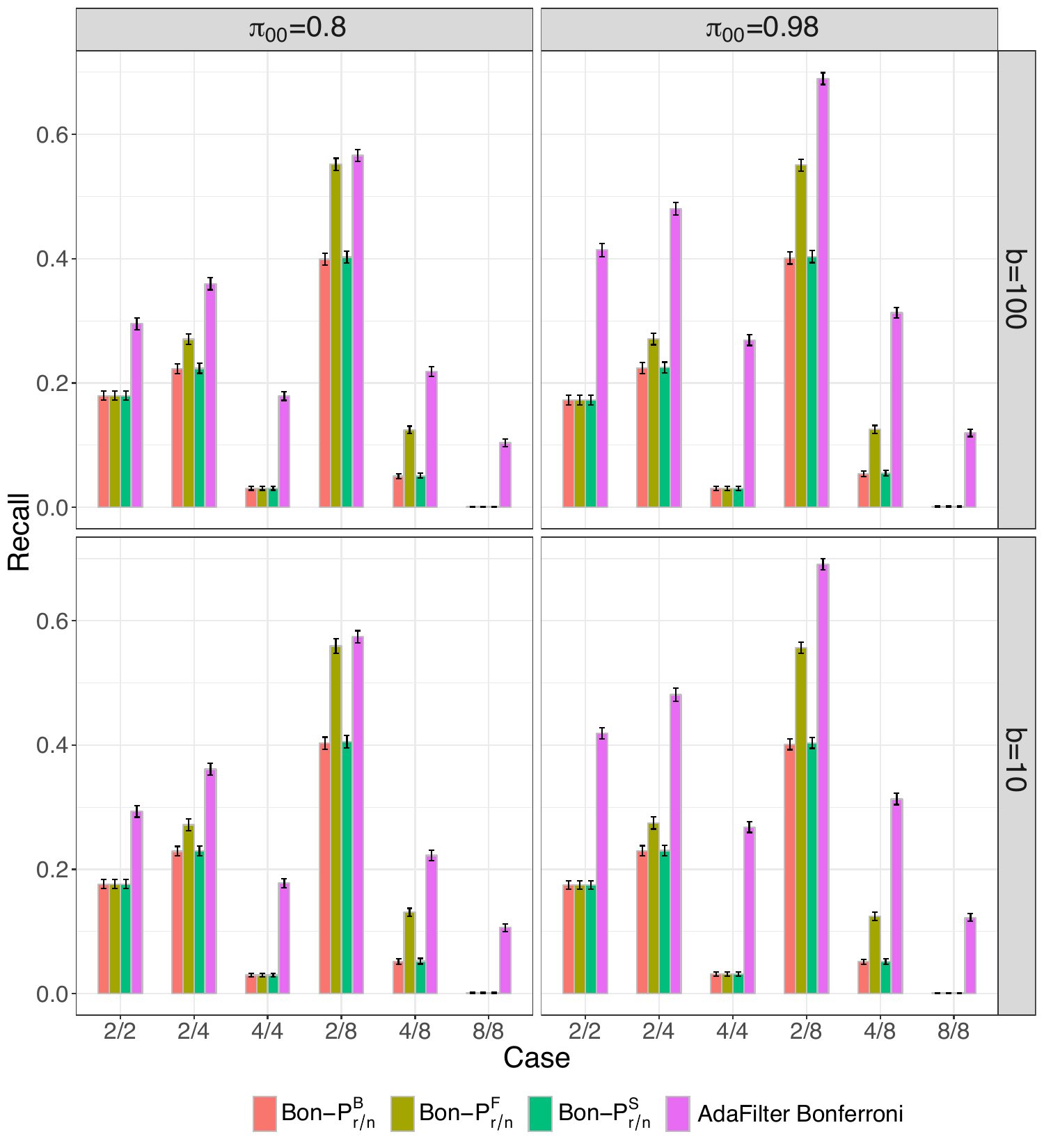}
\caption{Comparison of power (recall or sensitivity) when PFER is controlled at $\alpha = 1$. The error bars are the $95\%$ CI of the recall
from $B = 100$ experiments. The number of blocks $b = 100$ is the weak dependence scenario and $b = 10$ is the strong dependence scenario.}
\label{fig:power_two}
\end{figure}

\begin{figure}
  \centering
\includegraphics[width = 0.85\textwidth]{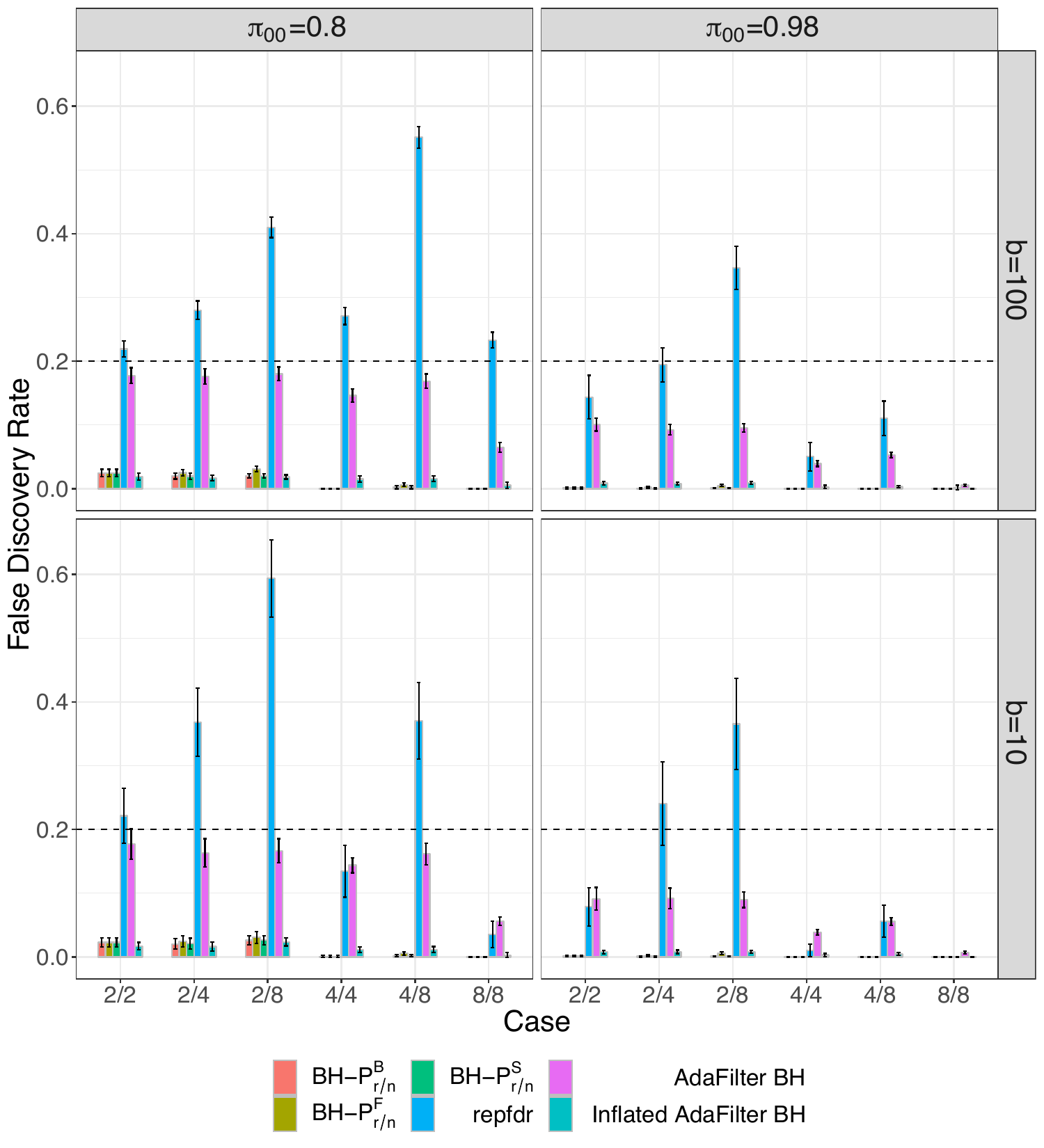}
\caption{Comparison of false discoveries rate $E(V/R)$ (FDR). The 
dotted line indicates the nominal level $\alpha = 0.2$. The error bars are the $95\%$ 
CI of the FDR.}
\label{fig:fdr_two_bh}
\end{figure}

\begin{figure}
  \centering
\includegraphics[width = 0.85\textwidth]{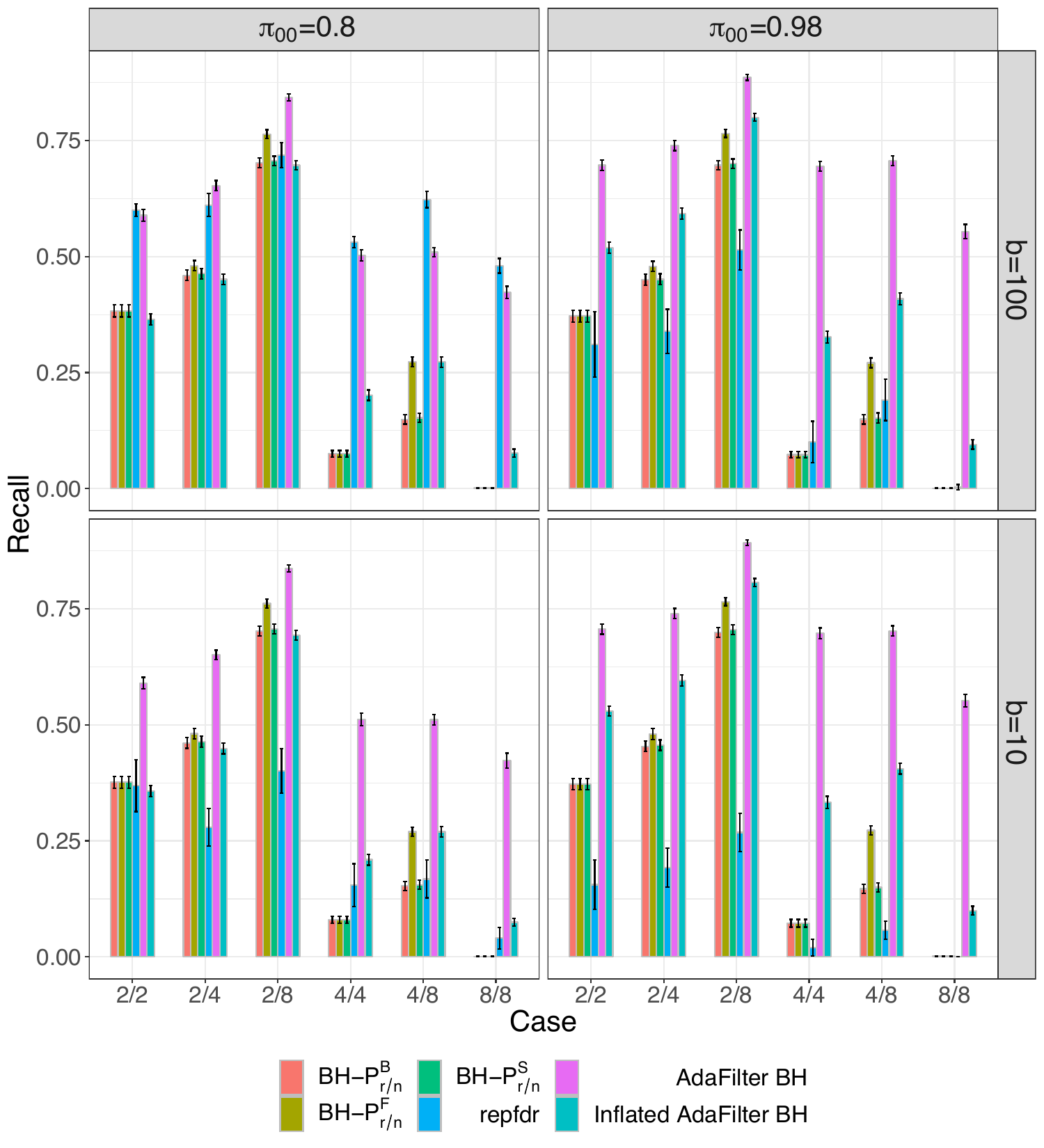}
\caption{Comparison of power (recall or sensitivity) when FDR is controlled at level $\alpha = 0.2$. The error bars are the $95\%$ CI of the recall 
from $B = 100$ experiments.}
\label{fig:power_two_bh}
\end{figure}


\end{document}